\documentclass[fleqn,usenatbib]{mnras}

\usepackage{newtxtext,newtxmath}

\usepackage[T1]{fontenc}

\DeclareRobustCommand{\VAN}[3]{#2}
\let\VANthebibliography\thebibliography
\def\thebibliography{\DeclareRobustCommand{\VAN}[3]{##3}\VANthebibliography}

\usepackage{graphicx}	
\usepackage{amsmath}	

\usepackage{colortbl}   
\usepackage{graphicx}   
\usepackage{float}

\newbox\grsign \setbox\grsign=\hbox{$>$} \newdimen\grdimen \grdimen=\ht\grsign
\newbox\simlessbox \newbox\simgreatbox
\setbox\simgreatbox=\hbox{\raise.5ex\hbox{$>$}\llap
     {\lower.5ex\hbox{$\sim$}}}\ht1=\grdimen\dp1=0pt
\setbox\simlessbox=\hbox{\raise.5ex\hbox{$<$}\llap
     {\lower.5ex\hbox{$\sim$}}}\ht2=\grdimen\dp2=0pt

\newbox\simppropto
\setbox\simppropto=\hbox{\raise.5ex\hbox{$\sim$}\llap
     {\lower.5ex\hbox{$\propto$}}}\ht2=\grdimen\dp2=0pt

\usepackage{tabularx}

\title[Measuring and modelling the Splash]{Measuring and modelling the Splash with APOGEE/Gaia and ARTEMIS}

\author[S. S. Kisku]{Shobhit Kisku$^{1}$,
Ricardo P. Schiavon$^{1}$\thanks{E-mail: R.P.Schiavon@ljmu.ac.uk},
Andreea S. Font$^{1}$,
Andrew Mason$^{1}$,
\newauthor Danny Horta$^{1,2,3}$,
Dominic J. Taylor$^{1,4}$,
Andrea Sante$^{1}$,
Jos\'e G. Fern\'andez-Trincado$^{5}$,
\newauthor Timothy C. Beers$^{6}$
\\
\\
$^{1}$Astrophysics Research Institute, Liverpool John Moores University, 146 Brownlow Hill, Liverpool L3 5RF, UK\\
$^{2}$Center for Computational Astrophysics, Flatiron Institute, 162 5th Ave., New York, NY 10010, USA\\
$^{3}$Institute for Astronomy, University of Edinburgh, Royal Observatory Edinburgh, Blackford Hill, Edinburgh EH93HJ, UK \\
$^{4}$Centre for Extragalactic Astronomy, Department of Physics, Durham University, South Road, Durham DH1 3LE, UK \\
$^{5}$ Instituto de Astronom\'ia, Universidad Cat\'olica del Norte, Av. Angamos 0610, Antofagasta, Chile\\
$^{6}$ Department of Physics and Astronomy and JINA Center for the Evolution of the Elements (JINA-CEE), University of Notre Dame, Notre Dame, IN 46556, USA \\
}
\date{Accepted XXX. Received YYY; in original form ZZZ}

\pubyear{2022}

\begin{document}
\label{firstpage}
\pagerange{\pageref{firstpage}--\pageref{lastpage}}
\maketitle

\begin{abstract}

Using combined data from SDSS-IV/APOGEE and Gaia, we study the
chemo-dynamical properties of the {\it Splash} population in comparison
with those of the high-$\upalpha$ disc. We investigate a wide range of
abundance ratios, finding that the {\it Splash} differs from the high-$\upalpha$
disc overall.  However, these differences result from a smooth
variation of chemical compositions as a function of orbital properties.  The {\it Splash} occupies the high-$\upalpha$, high-[Al,K/Fe], and low-[Mn/Fe]
end of the high-$\upalpha$ disk population.  In agreement with previous
studies, we find that {\it Splash} stars are distributed over large heights from the Galactic mid-plane. To further elucidate the
relation between the {\it Splash} and the high-$\upalpha$ disk, we turn to
simulations. Using a sample of Milky Way-like galaxies with and
without major accretion events from the ARTEMIS simulations, we
find that {\it Splash}-like populations are ubiquitous, though not always
resulting from major mergers. Lower mass progenitors can also
generate {\it Splash}-like features, as long as they are on retrograde
orbits. Moreover, we find a strong correlation between the mass
fraction of {\it Splash} stars and the fraction of retrograde accreted
stars in the disk. Some galaxies with minor (retrograde) mergers
contain more pronounced {\it Splash} populations than others with major,
but prograde, mergers. For stars in the high-$\upalpha$ disks, we also find a decrease in the [$\upalpha$/Fe] with increasing orbital angular momentum. This trend is found in hosts with both major or minor mergers. Our results suggest that a number of relatively low-mass
mergers on retrograde orbits could result in populations that are
qualitatively similar to the {\it Splash}.
\end{abstract} 





\section{Introduction}
\label{intro}

The field of galactic archaeology has seen a lot of major advances recently with large-scale surveys such as the Sloan Digital Sky Survey (SDSS, \citealp{blanton2017}), the Apache Point Galactic Evolution Experiment (APOGEE, \citealp{majewski2017}), and {\it Gaia} (\citealp{gaia2018}) leading to the discovery of many substructures in the Milky Way (MW) halo (\citealp{belokurov2006, schiavon2017, helmi2018, haywood2018, belokurov2018, mackereth2019, naidu2020, horta2021}).  
Among these discoveries was the identification of the {\it Splash}, 
a retrograde stellar population thought to originate from the heating of the MW disc by the merger with a massive satellite galaxy dubbed the Gaia-Enceladus/Sausage (GE/S) \citep{belokurov2018,helmi2018}. Further studies have revealed that the chemistry of the {\it Splash} is similar to that of the high-$\upalpha$ disc, but its kinematics are more halo-like (\citealp{bonaca2017, dimatteo2019, belokurov2020}).  This is consistent with the properties of the {\it in situ} halo, a component predicted by simulations \citep{zolotov2009,mccarthy2012, cooper2015} and thought to be the product of the heating of the MW disc by mergers.  

By combining data from a variety of sources, \citet{belokurov2020} studied the chemo-dynamic properties of the thin and thick disc and the halo within the solar neighbourhood.  They associate the Splash to a large population of relatively metal-rich stars ([Fe/H]$>-0.7$), which overlaps with the thick disc according to multiple properties, in particular in chemical compositions, as measured by the [$\upalpha$/Fe] ratios, and in velocity dispersion.  Therefore, they hypothesise that the origin of the Splash is linked to the old thick disc. However, while the chemical properties of Splash stars are similar to those of their thick disc counterparts, their orbits are far more eccentric (\citealp{mackereth2019, belokurov2020,ortigoza2023}). Moreover, Splash stars are found to be older than the average thick disc stars, yet slightly younger than the majority of known accreted structures (e.g., GE/S). Specifically, the age distribution of the Splash is truncated around the ages of the youngest stars associated with GE/S \citep{helmi2018, belokurov2018}).  
\citet{belokurov2020}.  Similarly, on the basis of {\it Gaia} data, \citet{gallart2019} show that the disc of the MW experienced a second peak of star formation, which they use to date the merger event at around $9.5$~Gyr ago.  
On the other hand, based on ages for a sample of sub-giants from the LAMOST survey (\citealp{cui2012}), \citet{xiang2022} find that the thick disc population only has a single peak in the star formation at $11.2$~Gyr ago, suggesting that the merger with GE/S was completed around that time \citep[but see][]{horta2024}. 

These findings suggest that the Splash is caused by the merger with GE/S, which heated the disc of the MW and ejected disc stars into the halo.  This implies that the Splash may contain the oldest stellar population of the MW disc, so its study should provide crucial information about the early stages of Galactic disc formation.  
Further studies of the early MW disc by \citet{belokurov2022} reveal a more metal-poor population of {\it in situ} stars, which they name {\it Aurora}, with kinematics and chemistry that differ from those of the Splash.  These differences are more clear at [Fe/H] around $-0.9$. {\it Aurora} shows a larger scatter in many abundances  as well as in its tangential velocity distribution, than what is seen in the Splash, cementing it as a different population. 

Along with large-scale surveys we have also seen improvements in numerical simulations to the point where high-resolution MW-like zoom-ins are a good match to detailed properties of the Milky Way.  This provides us with a much larger sample of galaxies which can be used to see where the MW fits in comparison, while also guiding the interpretations of observational data.  
Two recent simulation-based works that study the effects of mergers on the heating of the disc are \citet{grand2020} and \citet{dillamore2022}, which make use of simulations from the \texttt{AURIGA} project (\citealp{grand2017, grand2018}) and \texttt{ARTEMIS} (\citealp{font2020}), respectively.  
Both studies find that the age of the youngest Splash stars coincides with the end of a GE/S-like merger, in line with the observational results in the MW from 
\citet{belokurov2020}.  They also find a relation between the fraction of Splash stars and the mass of the Most Massive Accreted Progenitor (MMAP).

\citet{grand2020} identify a burst of star formation in the host galaxies right after the end of the MMAP merger event, making a prediction for the existence of this starburst in the MW. Similar starbursts are also frequent in the simulations studied by \citet{dillamore2022}.  
These authors also examine the orbital properties of the Splash population (which they define as {\it in situ} stars on retrograde orbits at present-time) and find that in most galaxies there is noticeable change in their orbits at the end of the MMAP accretion.  However, in a few cases, this change does not coincide with the end of the MMAP accretion, but rather with the end of another massive accretion event. These results provide good evidence that Splash-like populations may not be necessarily related to MMAP events.

An alternative explanation for the Splash is given by \citet{amarante2020}, who use a hydrodynamical simulation of an isolated galaxy that develops clumps in high-density regions of the disc and also produces a disc which displays an $\upalpha$-bimodality.  
These authors find that the clump scattering forms a metal-poor, low angular momentum, population, without the need for a merger.
Also recently, \citet{dillamore2023a} examined the high density of stars located in the same region as the Splash in the energy -- angular momentum plane and proposed that such a population may be caused by the effects of bar resonance.

In this paper, we study the chemo-dynamical properties of the Splash and contrast it with those of the high-$\upalpha$ disc. The paper has two parts: in the first, we present a detailed investigation of a wide range abundance ratios in the two populations, in various spatial and kinematical ranges. In the second part, we use simulated MW-type galaxies from the \texttt{ARTEMIS} suite to investigate the frequency of Splash features and their correlation with the type of accretion histories (i.e., major events such as GE/S-like, or more minor events). 
In Section~\ref{sec:data} we present the observational data and motivate our sample selection.  Our methods for the analysis of the chemical properties of the Splash and disc populations are discussed in Section~\ref{sec:chemistry}, and our observational results are presented in Section~\ref{sec:result}.  
In Section~\ref{artemis_sample} we study the chemo-dynamical properties of Splash-like populations and high-$\upalpha$ discs in {\texttt ARTEMIS} simulations, focusing on comparing galaxies which have undergone a GE/S-mass merger (henceforth `MW-GES') with those that only experienced minor accretions, (henceforth,`MW-MA') systems. Section~\ref{sec:discussion} includes a discussion and a summary of our findings.

\section{Data and Sample}
\label{sec:data}

We make use of the 17th data release (DR17, \citealp{abdurrouf2022}) from the SDSS-IV/APOGEE-2 survey (\citealp{blanton2017, majewski2017}), along with distances and velocities based on {\it Gaia} EDR3 (\citealp{gaia2021}) and \texttt{astroNN}\footnote{https://github.com/henrysky/astroNN} (\citealp{leung2019}).  
APOGEE is a near infra-red (H-band) spectroscopic survey based on both the northern and southern hemispheres.  With the use of the 2.5m Sloan Foundation Telescope at Apache Point Observatory (APO, \citealp{gunn2006}) in the United States (APOGEE-2N) and the 2.5m du Pont Telescope (\citealp{bowen1973}) at Las Campanas Observatory in Chile (APOGEE-2S), APOGEE is able to achieve full sky coverage with data for over 650,000 unique targets.  These telescopes are fitted with twin high-resolution (R $\sim$ 22,500) multi-fibre spectrographs (\citealp{wilson2019}) providing abundances for over 20 different elements.  Abundances are derived from the automatic analysis of spectra using the ASPCAP pipeline (\citealp{garcia2016,jonsson2020}).  For further details and information on data reduction and selection criteria, see \citet{nidever2015, holtzman2015, holtzman2018, zasowski2017, beaton2021, santana2021}.  

In addition to the above data, orbital information was calculated as follows.  First, 6D phase-space information based on Gaia eDR3 astrometry and proper motions and APOGEE DR17 radial velocities was converted to Galactocentric cylindrical coordinates, adopting a solar velocity vector given by [U$_\odot$, V$_\odot$, W$_\odot$] = [–11.1, 248.0, 8.5]~km~s$^{-1}$. The latter was based on the Sgr~A$^\star$ proper motion from \cite{reid2020} and the local standard of rest by \cite{Schonrich2010}. For the position of the Sun, we adopted a distance of 8.178~kpc from the Galactic centre \citep{grav2019} and a distance of 20~pc from the Galactic mid-plane \citep{bennett2019}.  Finally, orbital parameters were computed using the publicly available {\texttt{galpy}} software, \citep{bovy2015,mackereth2018}.  We adopted the fast parameter estimation technique, which uses the St\"ackel fudge method for estimating action-angle coordinates in axisymmetric potentials \citep{binney2012}. The Galactic potential by \cite{mcmillan2017} was assumed.

\subsection{Sample selection}
\label{sample}

To obtain the best abundances, we limit our parent sample to G-K giant stars.  In addition, we only select the primary survey targets by setting the $\texttt{EXTRATARG}=0$, and make some quality cuts to obtain a clean sample.  Our selection criteria can be summarized as follows. 
\begin{itemize}
    \item ${\rm d_\odot}<3$ kpc 
    \item $1<\rm{log}{\it g}<3$
    \item 4000 K$<T_{\rm{eff}}$ $<6000$ K
    \item $\rm{SNR}>100$ 
    \item $\rm{d_{err}/d}<0.2$ 
    \item $\texttt{ASPCAPFLAG}=0$
    \item $\texttt{EXTRATARG}=0$
    \item $\rm{[X/Fe]}>-10$ (X being each element used in the analysis)
\end{itemize}

The $\rm{[X/Fe]}$ criterion aims to remove stars for which a given elemental abundance could not be delivered by ASPCAP (listed in the catalogue as equal to -9999).  We also remove all stars contained in the DR17 Value Added Catalogue of Galactic globular cluster stars \citep{Schiavon2024}. 
For the reasons described in Section~\ref{logg}, we restrict our sample to the solar neighbourhood by limiting the heliocentric distances to 3~kpc.  

\begin{figure}
    \includegraphics[width=\columnwidth]{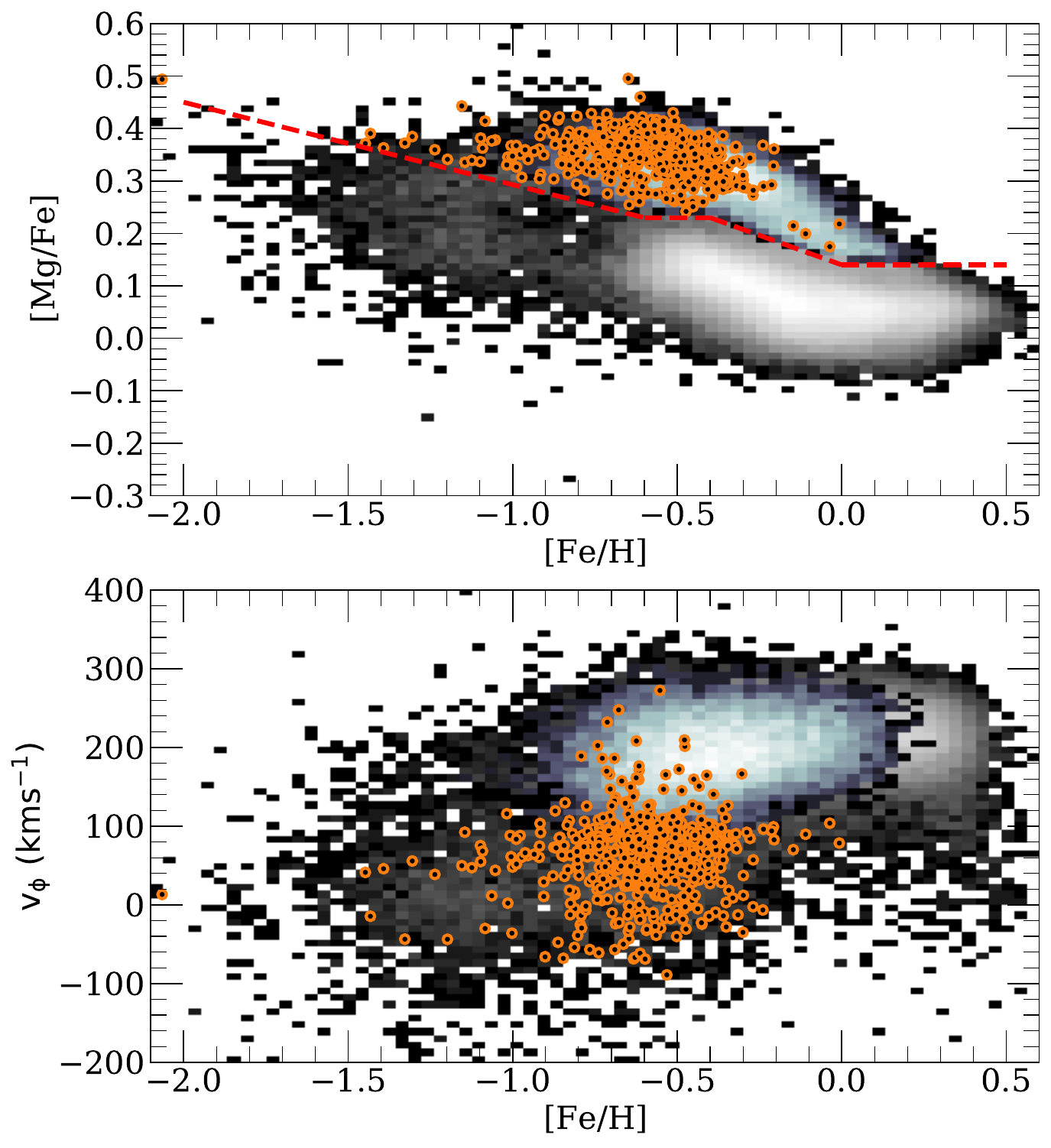}
    \caption{[Mg/Fe]---[Fe/H] ({\it top}) and $v_{\phi}$ -- [Fe/H] ({\it bottom}) distributions of the star samples. The clean APOGEE data are shown in greyscale as 2D histograms.  Over-plotted are the high-$\upalpha$ disc, shown as blue histograms, and the Splash, as orange dots, both limited to the solar neighbourhood.  A cut is made in the [Mg/Fe]---[Fe/H] plane to select the high-$\upalpha$ disc stars, as indicated by the red dashed line in the top panel. Our selection of the Splash places it in the same locus in the $v_{\phi}$ -- [Fe/H] plane as in \citet{belokurov2020} and \citet{belokurov2022}.}
    \label{mgfe}
\end{figure}

Since we are mainly concerned with a comparison between the Splash and the high-$\upalpha$ disc, we further adopt a chemical composition selection\footnote{We note that we experimented with making different selections for the Splash and a comparable selection of the high-$\upalpha$ disc, independent of chemistry, which would be better suited for a chemical comparison.  
However, this led to a large contamination with GE/S stars in the Splash sample. We ultimately chose to impose a chemical cut to minimize contaminants.  After including a chemical cut, as shown in the top panel of Fig.~\ref{mgfe}, we found that the simplest additional cut was to separate the Splash and high-$\upalpha$ disc according to eccentricity.  
As shown in the bottom panel of Fig.~\ref{mgfe}, these cuts select a Splash population that lies in the same region as highlighted by \citet{belokurov2020}.} to isolate a clean high-$\upalpha$ disc sample.  
The latter is made on the basis of the [Mg/Fe]--[Fe/H] plane, as shown in Fig.~\ref{mgfe} (note that this figure includes corrections described in Section~\ref{motive}), selecting those stars that lie above the red dashed line as high-$\upalpha$ members.  

\begin{figure}
    \centering
    \includegraphics[width=\columnwidth]{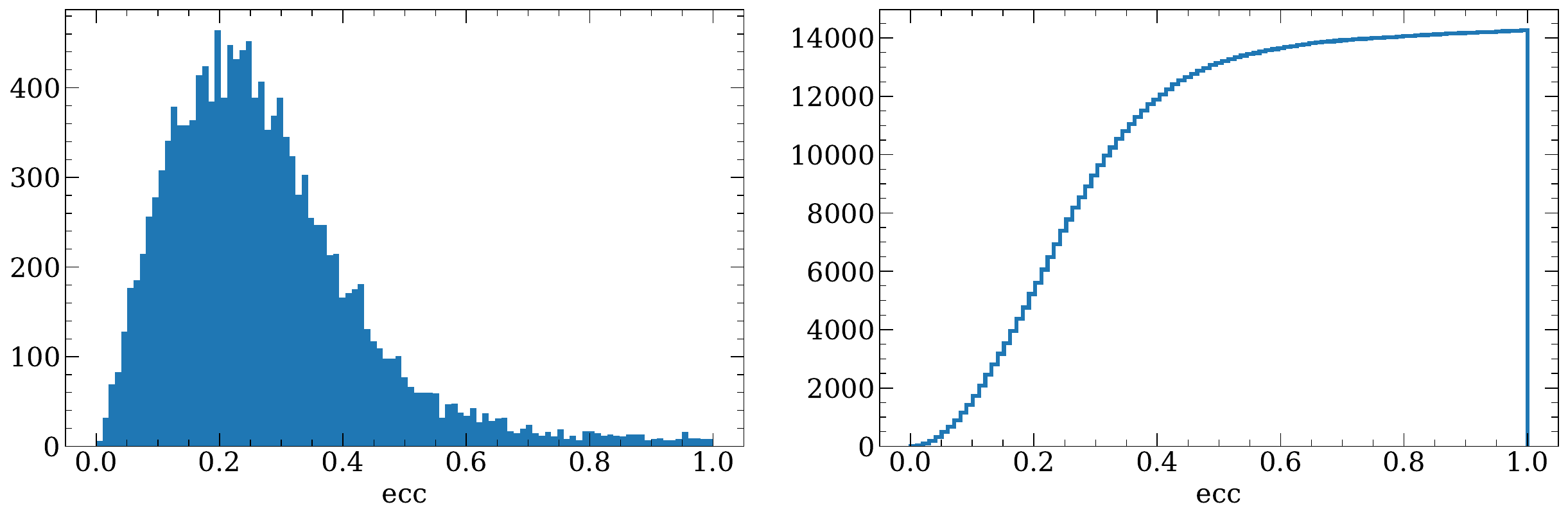}
    \caption{The eccentricity distribution for the high-$\upalpha$ population.  We select the Splash to be in the upper tail end of this distribution where there is a significant change in the gradient, i.e., at $e>0.6$.}
    \label{ecc_hist}
\end{figure}

A simple eccentricity cut is then made to the high-$\upalpha$ parent sample in order to separate the disc and the Splash populations. Specifically, stars with $e>0.6$ are considered to belong to the Splash. We choose this specific eccentricity cut-off because this is where we see a turn-over in the gradient of the eccentricity distribution in the high-$\upalpha$ disc, turning from a normal Gaussian tail into a flat extended tail (see Fig.~\ref{ecc_hist}).

\vspace{8pt}
To summarize, the selection criteria for the high-$\upalpha$ disc stars are: 
\begin{itemize}
    \item The parent sample defined above
    \item high $\rm{[Mg/Fe]}$ (above the red dashed line in the top panel of Fig.~\ref{mgfe})
    \item \bf{$e<0.6$}
\end{itemize}

\noindent and for the Splash: 
\begin{itemize}
    \item The parent sample defined above
    \item high $\rm{[Mg/Fe]}$ (above the red dashed line in the top panel of Fig.~\ref{mgfe})
    \item $e>0.6$.
\end{itemize}

The high-$\upalpha$ parent sample consists of 14,258 stars, and the Splash sample contains 626 stars.

\subsection{Motivations for the selection criteria}
\label{motive}

\subsubsection{The APOGEE selection function}

\begin{figure}
    \includegraphics[width=\columnwidth]{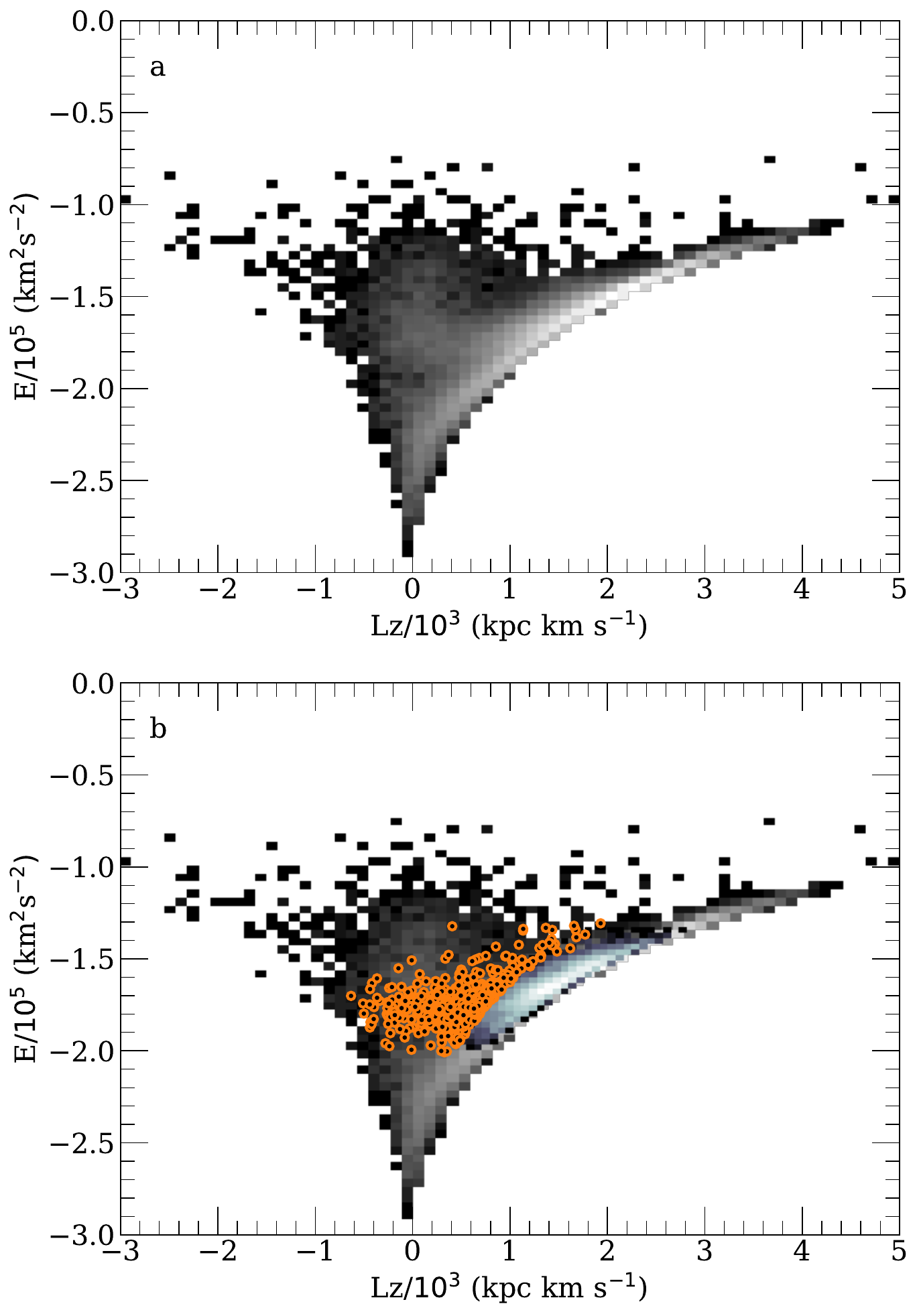}
    \caption{{\it Top:} Energy-angular momentum of stars in the APOGEE catalogue, shown as a 2D histogram, using the selection criteria described in Section \ref{sample}.  Positive Lz represents prograde motion.  The tuning fork feature is visible with a gap at $\sim(0, -2.0\times10^5)$.  {\it Bottom:} Same as in the top panel, but highlighting the Splash (orange) and high-$\upalpha$ disc (blue) selections in the solar neighbourhood.}
    \label{elz}
\end{figure}

By limiting our sample to the solar neighbourhood, we may miss important information about the nature of the Splash as seen on the Galactic level.  However, over large distances, the APOGEE's selection function introduces additional biases.  For example, in the $E-L_z$ plane shown in Fig.~\ref{elz}, there is a conspicuous branch that extends from the disc locus around $E\sim-1.8\times10^5$~km$^2$ s$^{-2}$. \citet{horta2021} associated this branch with the Splash population.  However, as shown in \citet{lane2022}, this overdense population may be a result of the APOGEE selection function. By limiting our analysis to the solar neighbourhood, we minimize the potential effects of the selection function on our results.

\begin{figure}
    \centering
    \includegraphics[width=\columnwidth]{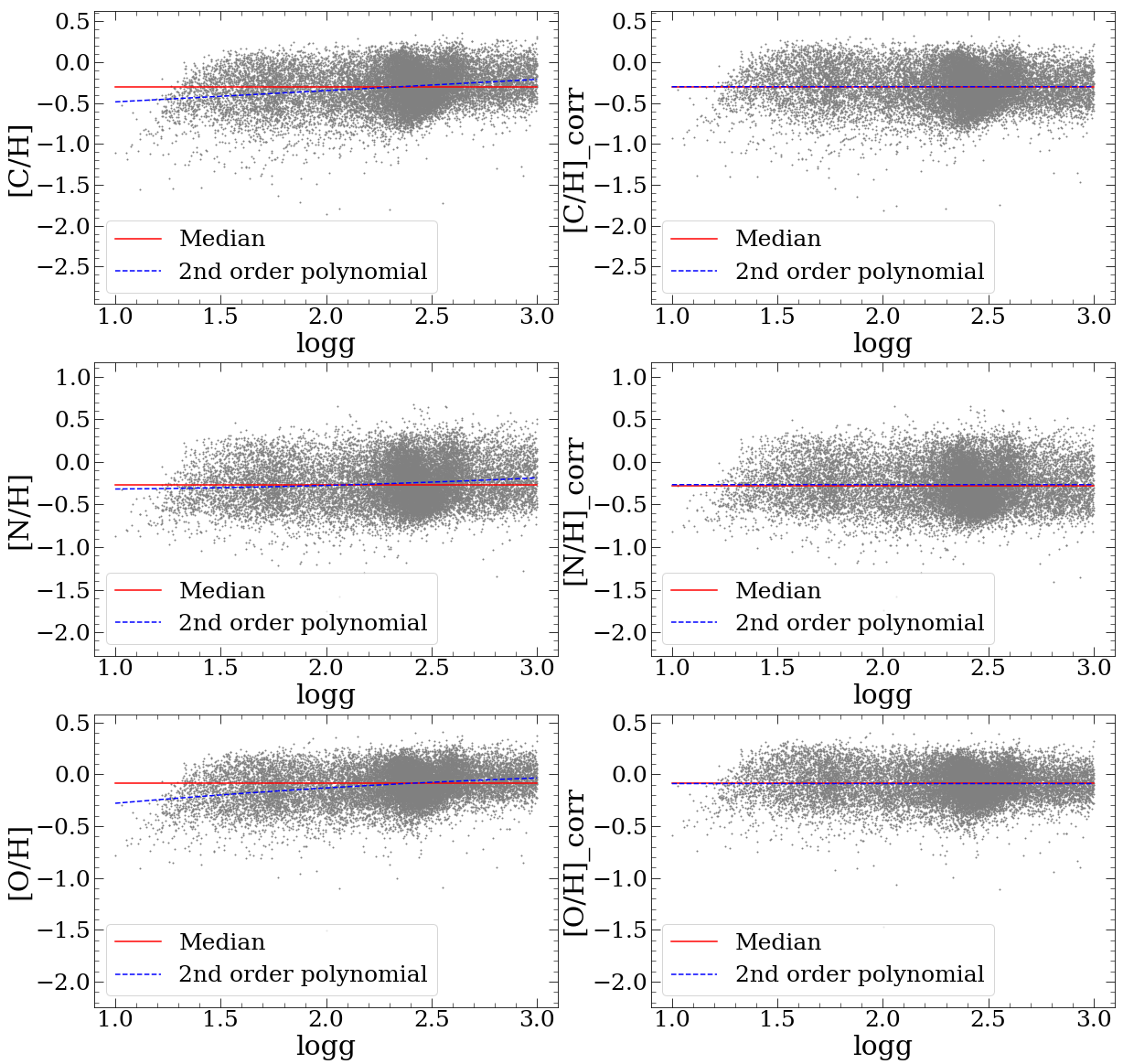}
    \caption{Example visualisation of the correction for abundance systematics on the [X/H] - log$g$ plane.  The left and right panels show the uncorrected and corrected abundances, respectively.  Grey points show the high-$\upalpha$ disc stars at the solar radius, while the red full and blue dashed lines show the medians of the distributions and the $2^{\rm nd}$ order polynomials, respectively.}
    \label{loggcorrection}
\end{figure}

\begin{figure*}
    \includegraphics[width=\textwidth]{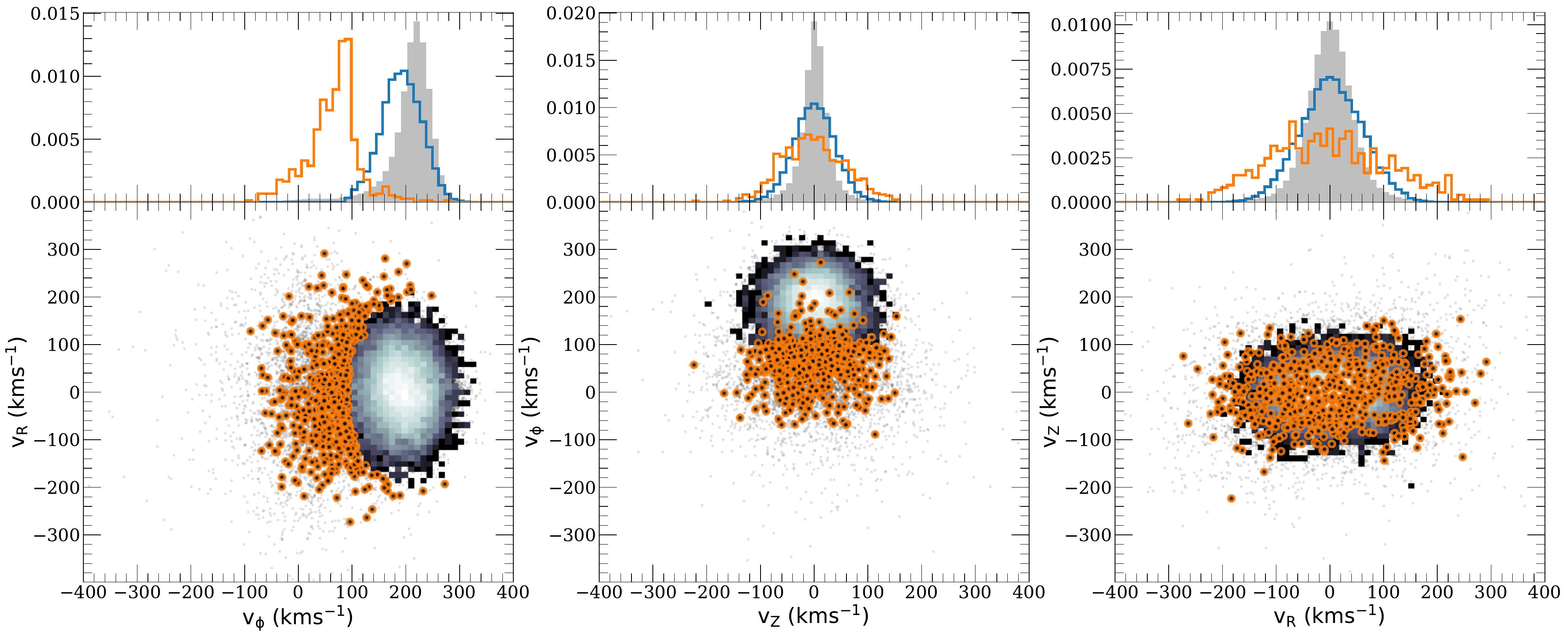}
    \caption{Bottom/top panels show the 2D / 1D velocity distributions of the Splash (orange dots/orange histograms), high-$\upalpha$ disc (2D / blue histograms), and the full clean APOGEE sample (black dots / filled grey histograms), respectively.  The 1D histograms share the same x-axis as the plots below.  The left panels indicate a clear difference between the Splash and the high-$\upalpha$ disc in $\rm{v_\phi}$.  For $\rm{v_Z}$ and $\rm{v_R}$, the 2D distributions for our two samples are more similar, however, the 1D histograms show that the Splash has slightly more extended distributions in both velocities.  These properties result from the selection criteria adopted to define the Splash.}
    \label{velocity}
\end{figure*}

\begin{figure*}
    \centering
    \includegraphics[width=\textwidth]{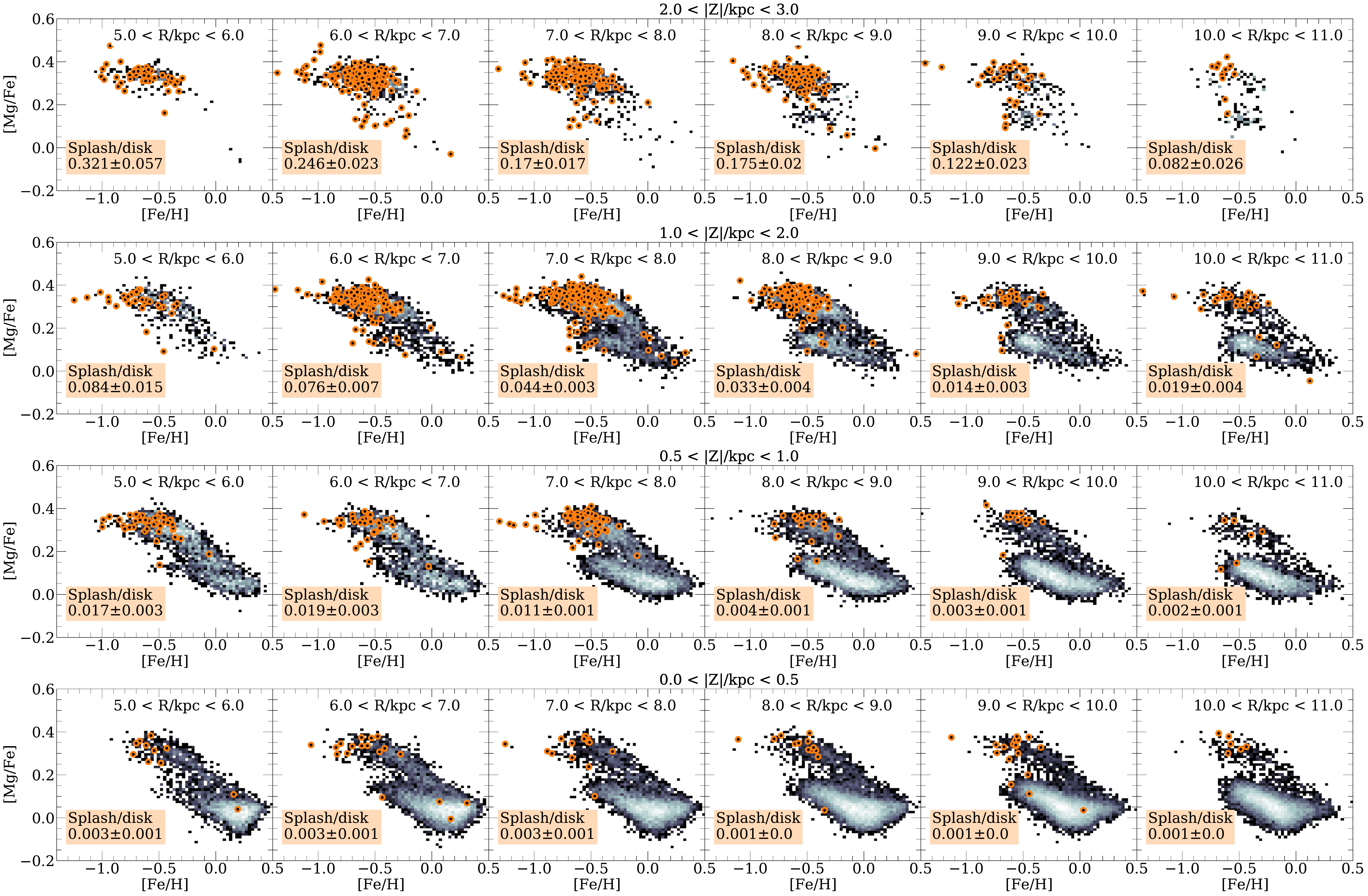}
    \caption{Distribution of the Splash (orange dots) and high-$\upalpha$ disc (2D histograms) in the [Mg/Fe]--[Fe/H] plane as a function of Galactocentric radii ($R$/kpc) and absolute vertical height ($|Z|$/kpc).  We also show the ratio of Splash to high-$\upalpha$ disc in the metallicity range $-1.1<$[Fe/H]$<-0.3$, marked by black dashed lines, the errors for which are taken to be the standard error for the number of stars.  The Splash fraction increases with increasing height ($|Z|$), but decreases with increasing $R$.}
    \label{ratio}
\end{figure*}
\begin{figure}
    \centering
    \includegraphics[width=\columnwidth]{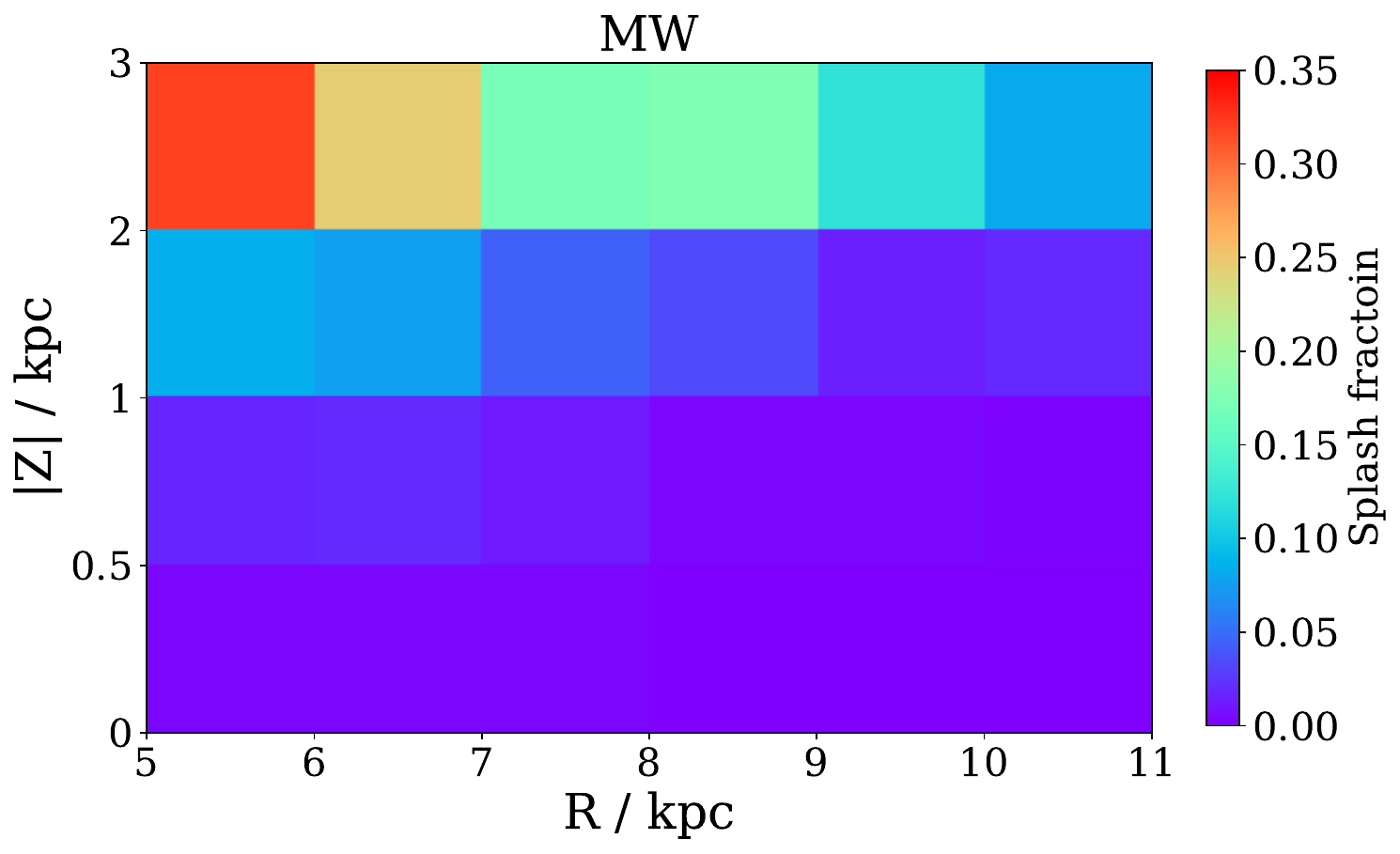}
    \caption{Confusion map showing the Splash fraction computed from each panel in Fig.~\ref{ratio}. This shows that the Splash fraction in the MW increases at higher values of $|Z|$ and at lower values of $R$.}
    \label{confusion}
\end{figure}

\subsubsection{Correction for abundance systematics}
\label{logg}
Recently, systematic trends in APOGEE abundances as a function $\log g$ have been reported by various groups \citep[e.g.,][]{eilers2022, Weinberg2022, horta2023}.  Such systematics must be corrected to enable unbiased comparisons between samples of stars with different $\log g$ distributions.  We perform this correction for each abundance individually adopting the following procedure.  First, we fit a second-order polynomial to the distribution of the data in various [X/H]--$\log g$ planes.  Second, we subtract the value of the polynomial from the median of the distribution to derive a residual correction to elemental abundances for a given value of $\log g$.  Third, we add this difference to the original abundance ratios. This is shown in Fig.~\ref{loggcorrection}.

\citet{eilers2022} show that there is a significant bias in the $\log g$ distribution of stars depending on the distance at which they are observed, with lower gravity stars being over-represented in more distant samples.  
By the same token, elemental abundances are known to vary as a function of position, so our correction for systematics as a function of $\log g$ could potentially erase such abundance gradients.  To remedy this potential issue, we restrict our sample to a relatively small spatial distribution.
Another way to minimise the effects of this systematic effect would be to select a narrower range of $\log g$, but that would reduce our sample to intolerably low numbers.

\subsection{Kinematics and Spatial Distributions}
\label{kinematics}

In this section, we examine the velocities and positions of Splash and high-$\upalpha$ disc stars, shown in Figs.~\ref{velocity} and \ref{ratio}, respectively.  
The bottom panels of Fig.~\ref{velocity} show the distribution of our sample stars in kinematic space (in cylindrical coordinates). As in previous plots, the orange dots represent the Splash stars and the 2D histograms indicate the number density distribution of the high-$\upalpha$ disc stars.  
We also include the entire clean APOGEE sample, defined in Section \ref{sample}, as small gray dots in the background.  The top panels display the distributions of the same sub-samples in the three components of the velocity vector, in the form of normalised 1D histograms, adopting a consistent colour scheme.

From the left and middle bottom panels it can be seen that the stars in the clean APOGEE sample extend to lower tangential velocities than those in the Splash, whereas the histogram in the top left panel shows that the majority of the APOGEE sample have much higher tangential velocities than the Splash.  From previous studies, we know that stars at such low tangential velocities are associated with the GE/S system, along with other less evolved accreted systems \citep{helmi2018, belokurov2018}.

Similarly, we find that the high-$\upalpha$ component is characterized by a slightly lower tangential velocity and a higher radial and vertical dispersion than the main APOGEE sample. This is also expected, given that a large part of the Galactic disc is comprised of a younger, low-$\upalpha$, thin disc population, which on average is more circular than the older, high-$\upalpha$ population.

Fig.~\ref{velocity} shows that a good qualitative agreement is achieved between the kinematic distribution of our Splash sample with that found in previous studies (e.g., \citealt{belokurov2018}). Moreover, it is clear that the two sub-samples display kinematical differences: the Splash stars have lower tangential velocity and a larger spread in both radial and vertical velocities than the high-$\upalpha$ disk stars. However, we note that the kinematical differences are expected, given that our Splash sample is selected in terms of its high eccentricity. In the following sections, we  investigate the chemical abundance differences between these two populations in order to determine whether the Splash is a distinct component from the high-$\alpha$ disc.

Next, we explore the spatial distribution of Splash stars, cast in terms of number counts as a function of position.  Because our data are not corrected for the APOGEE selection function, we must devise a measurement that is not sensitive to selection effects. 

To minimise these effects, we anchor our Splash star counts on the number of their high-$\upalpha$ disc counterparts, taking ratios within relatively small spatial bins. 
For this measurement only, we change the spatial cuts of our sample from a sphere around the Sun to an annulus of thickness 6~kpc centred on the Galactic centre, $5$~kpc $<R <11$~kpc and $|Z|<3$~kpc. 
We also include the low-$\upalpha$ region of the [Mg/Fe]--[Fe/H] plane for comparison with simulated galaxies presented in later sections.  
Fig.~\ref{ratio} shows the distribution of high-$\upalpha$ and Splash stars in the [Mg/Fe]--[Fe/H] plane, in bins of Galactocentric distance $(R)$ and vertical distance ($|Z|$) from the Galactic plane, in cylindrical coordinates.  We adopt the same colour scheme as used in  Figs.~\ref{mgfe}, \ref{elz} and \ref{velocity}.

As originally shown by \citet{hayden2015}, we find that, while the number of high-$\upalpha$ disc stars varies greatly with position, those stars lie roughly on the same $\upalpha$--Fe plane locus across the Galactic disc.  
The corresponding number ratio between the Splash and the high-$\upalpha$ disc stars (i.e., the Splash fraction), is shown in the insets of each panel.  A clear trend is seen between the Splash ratio and $|Z|$ and $R$, namely the further away from the plane, the larger the contribution of the Splash; also, the closer to the Galactic centre, the higher the contribution of the Splash. 

For another visualisation of this behaviour, we show in Fig.~\ref{confusion} the confusion map of the Splash fraction in the MW, colour-coded by the Splash fraction. This highlights the regions of highest density of Splash stars, notably at $R \simeq 5-7$~kpc and at  $|Z| \simeq 2-3$~kpc. 
However, we must keep in mind that our sample is restricted to the solar neighbourhood, and only reaches as close as 5~kpc from the Galactic centre. 

While the definitions of Splash may differ, our results are in good qualitative agreement with those obtained by \citealt{belokurov2020} (see their figures 5 -- 8). See also the simulation results of \citet{grand2020} (figures 4 and 5 in that study), which predict a similar spatial distribution of the Splash population relative to the disk.
In Section~\ref{sec:heatsim}, we  compare this confusion map with similar maps constructed from simulated Milky Way-mass galaxies, with and without a GE/S-like event, from the \texttt{ARTEMIS} suite.

\section{Chemistry of the Splash versus High-$\upalpha$ stars}
\label{sec:chemistry}

\begin{figure*}
    \centering
    \includegraphics[width=\textwidth]{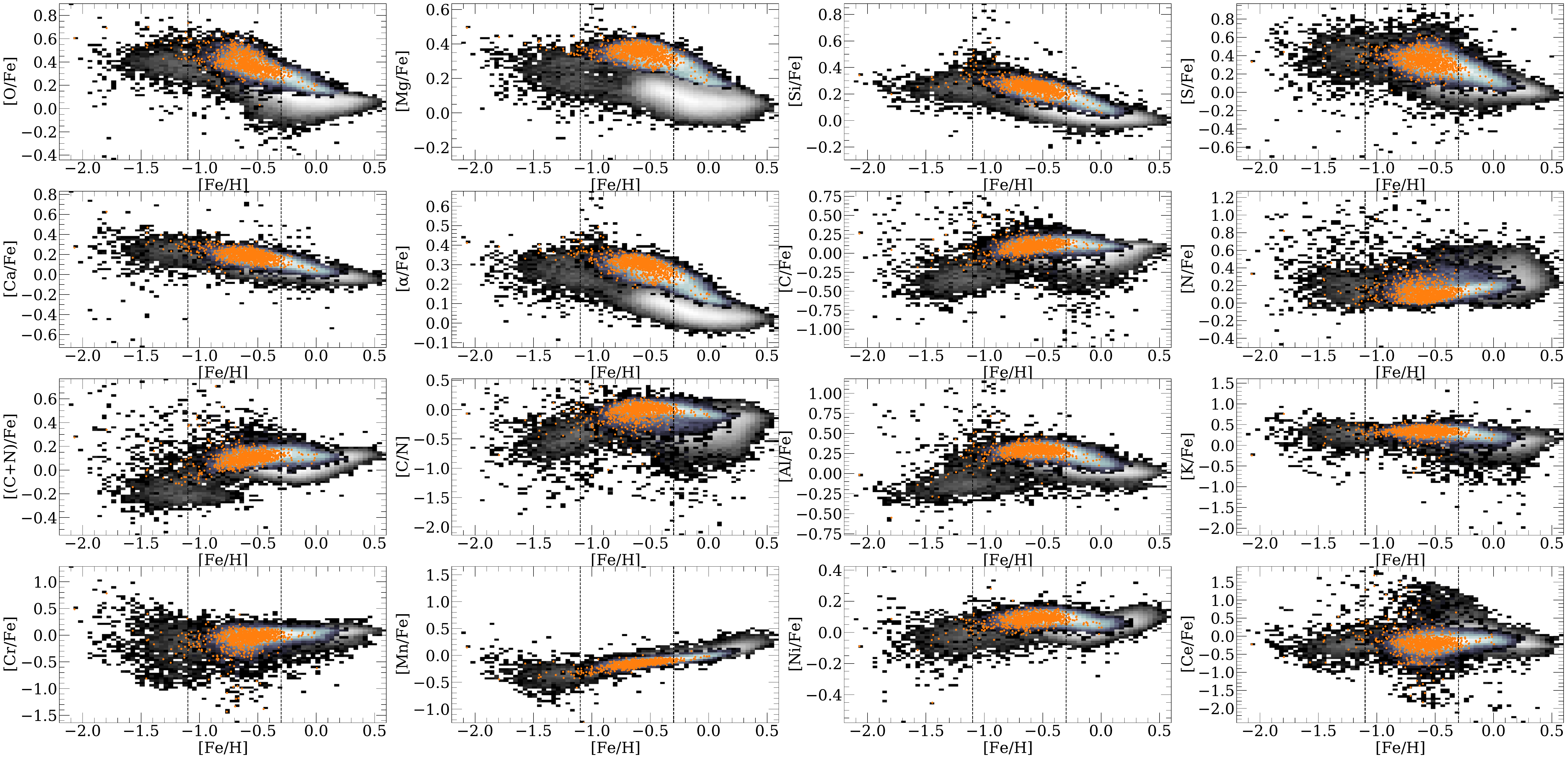}
    \includegraphics[width=\textwidth]{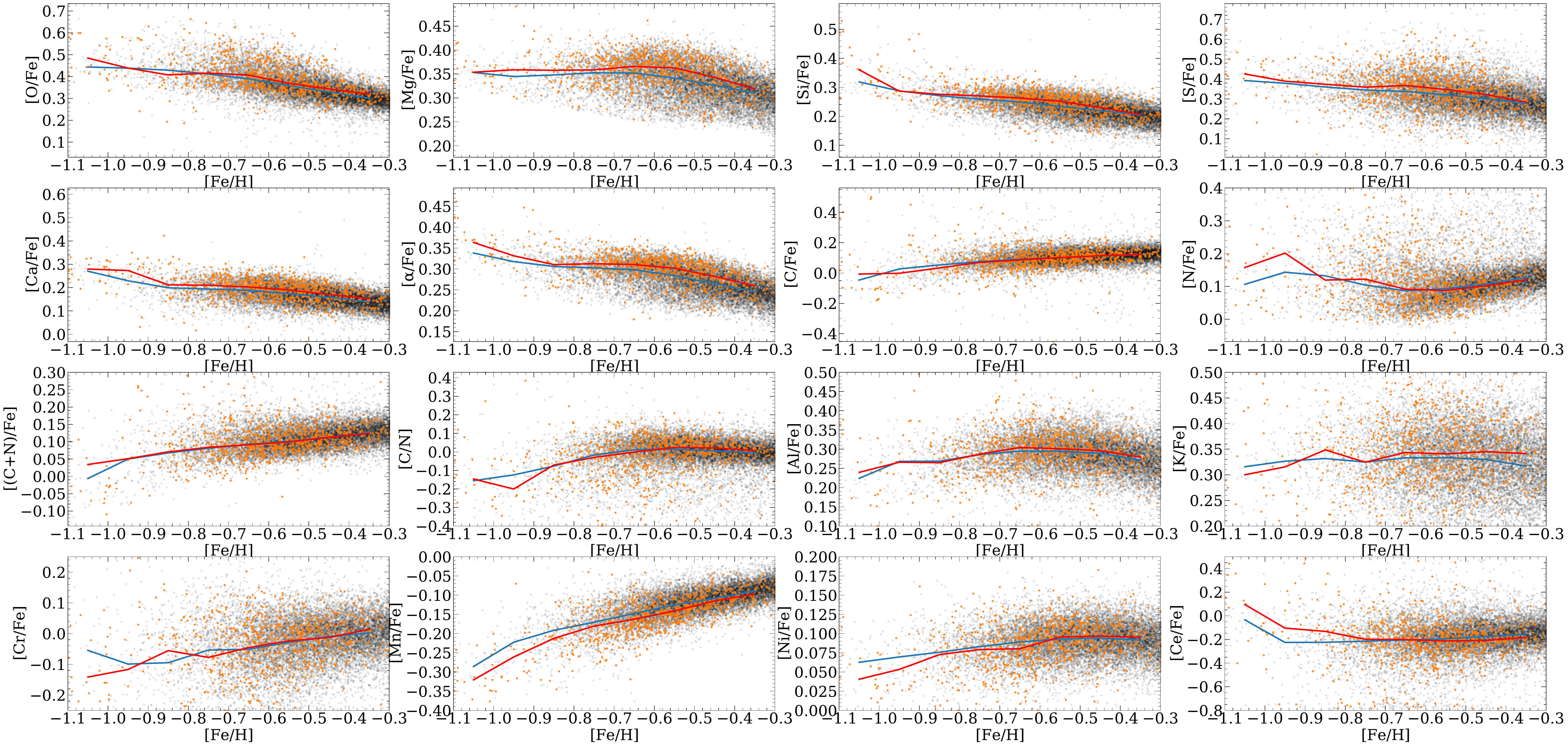}
    \caption{[X/Fe]--[Fe/H] planes for all abundance ratios analysed here.  The first 16 panels show the full metallicity range covered by APOGEE.  The grey 2D histograms show the fully cleaned APOGEE sample, with the high-$\upalpha$ disc stars shown as blue 2D histograms and the Splash stars as orange points.  The bottom 16 panels zoom into the metallicity range considered in this work, with Splash stars shown as orange points and high-$\upalpha$ disc stars as black points.  The running medians for both samples are shown in red for the Splash, and in blue for the high-$\upalpha$ disc stars, respectively.}
    \label{x-fe}
\end{figure*}

In this section we examine how the two populations compare in terms of their detailed abundance patterns.  Our goal is to establish whether these two populations differ chemically, and how that comparison leads up to a better understanding of the origin of the Splash population. In Fig.~\ref{x-fe} we display our populations in the [X/Fe]--[Fe/H] plane for all 16 elements. 
This figure has been split into two.  The top 16 panels show the 'zoomed out' version of the planes covering the entire APOGEE metallicity range.  The grey 2D histogram is the entire clean APOGEE sample, the high-$\upalpha$ population is shown as a blue 2D histogram, and the Splash population as the orange points. 
 In the bottom 16 panels we show the 'zoomed in' version, covering only the metallicity range for which we compare our Splash with the high-$\upalpha$ populations.  
 In the bottom panels, the high-$\upalpha$ population is shown as the black points and the Splash population as the orange points.  We also over-plot the running median of the two populations, with a bin size of 0.1~dex in metallicity, with blue for high-$\upalpha$ and red for Splash, respectively. 

In the following, we apply two different statistical methods to compare the chemistry of the Splash to the high-$\upalpha$ disc populations.  
The first method consists of splitting our samples into metallicity bins, [Fe/H], and calculating the average abundances for a selection of elements, similar to the one described in \citet{horta2023} (see Section \ref{danny}).  The second is similar to that of \citet{taylor2022}, where we compute a $\chi^2$ between the chemical abundances in the Splash and the high-$\upalpha$ disc, and compare it to the $\chi^2$ obtained by drawing 1,000 bootstrap samples of the high-$\upalpha$ disc, compared to the whole high-$\upalpha$ disc.  Each of the 1,000 bootstrap samples is precisely the same size as that of the Splash sample (see Section \ref{dom}).  

In the following, we compare the two populations only in the metallicity range $-1.1<$ [Fe/H] $<-0.3$.  The upper limit is due to the lack of more metal-rich Splash stars in our sample, which is the result of the eccentricity cut adopted to define the Splash population. This implicitly selects against metal-rich stars.  The lower limit is motivated by our aim to minimize the contamination by halo stars. 
We note that the selection cuts imposed in Sec.~\ref{sample} already achieves this to a great extent. The additional cut [Fe/H] $> -1.1$ imposed here aims to further minimize the contribution of the accreted halo. This limit is motivated by various results which show that [Fe/H] $\approx -1$ marks the onset of the disc (see, e.g., \citealt{chandra2024}). Nevertheless, we caution that near [Fe/H] $\approx -1$ it is generally difficult to separate the disc from the accreted halo due to the significant overlap of these components. 

The abundances adopted in our comparisons are the following: 1) $\upalpha$-abundance ratios: [O/Fe], [Mg/Fe], [Si/Fe], [S/Fe] and [Ca/Fe].  We also include a combination of these abundances by taking the inverse weighted average, which we label [$\upalpha$/Fe]; 2) light and odd-Z: [C/Fe], [N/Fe], [(C+N)/Fe], [Al/Fe] and [K/Fe]. For an indication of age, we also show [C/N]; 3) Iron-peak elements: [Cr/Fe], [Mn/Fe] and [Ni/Fe]; 4) {\it s}-process elements: [Ce/Fe].  
These abundances are discussed in Sections \ref{alpha}, \ref{lightoddz}, \ref{oddz}, \ref{ironpeak} and \ref{sprocess}, respectively.  Note that while we show the comparisons for elements that are a combination of others, i.e., [$\upalpha$/Fe], [(C+N)/Fe] and [C/N], we do not include them in the calculation used to obtain $\chi^2$.

\subsection{First Method: Abundance comparison}
\label{danny}
For this method, we follow a similar procedure to that used in \citet{horta2023}.  We first split our samples into bins of metallicity and compare the abundance ratios for all elements in each bin.  
Metallicity bins are chosen to have small enough widths to prevent any chemical evolution effects from affecting the sub-samples. 
The four [Fe/H] bins adopted are the following: [--1.1,--0.8], [--0.8,--0.6], [--0.6,--0.45], and [--0.45,--0.3]. Within each bin, median abundance ratios are computed for the two sub-samples. Errors are calculated through a method where we take 1,000 random bootstrap samples, with replacement, from our Splash and high-$\upalpha$ disc samples, individually, then calculate the standard deviation of the median values of the 1,000 random samples. 

\begin{figure*}
    \includegraphics[width=0.9\textwidth]{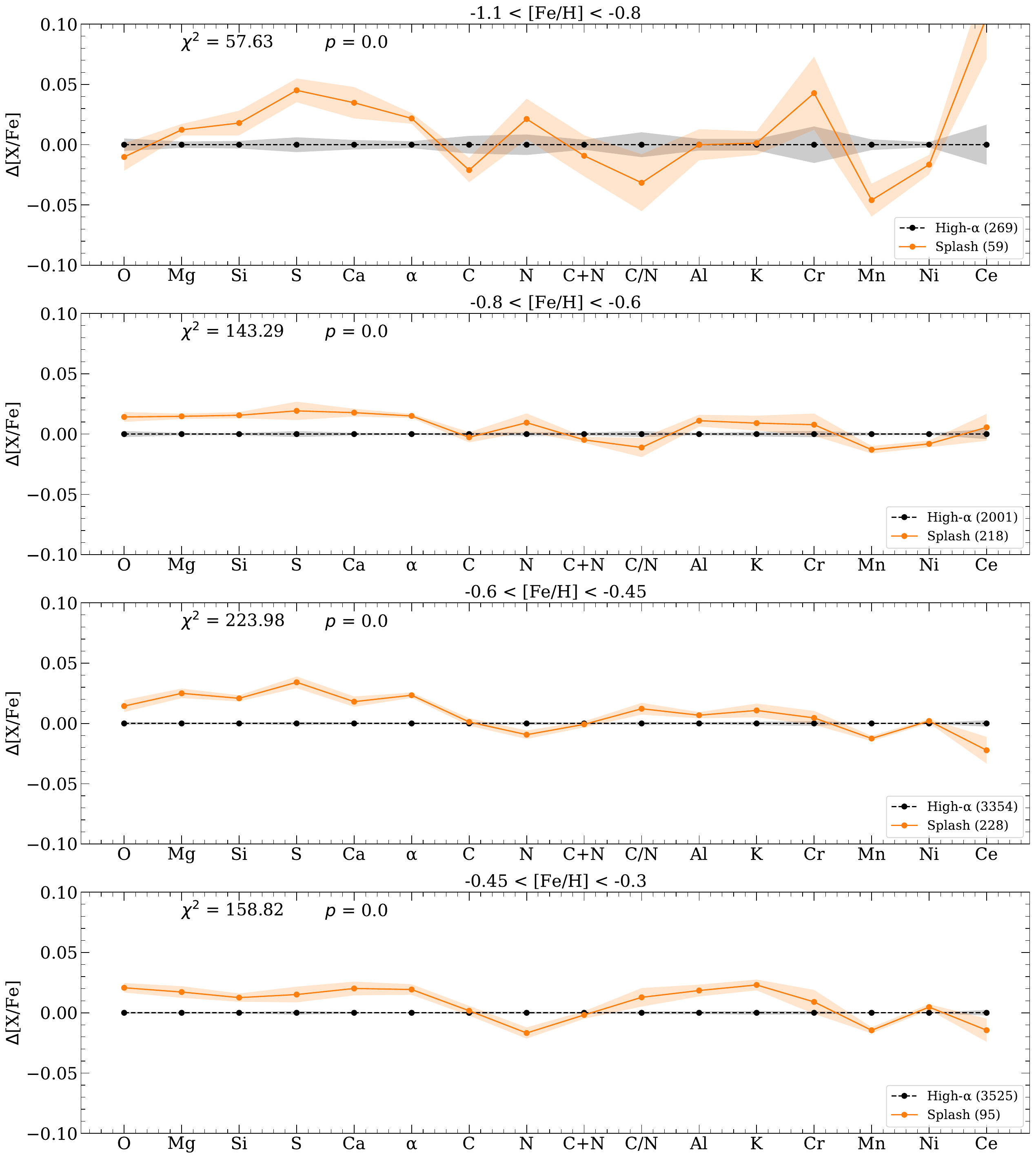}
    \caption{The chemical comparison plot between the Splash and the disc, the results for which are laid out in Section \ref{dannyresult}.  The points show the median values, orange for Splash and black for high-$\upalpha$ disc, with the medians of the high-$\upalpha$ subtracted from both, hence the black points all lie on zero.  The shaded regions are the respective 1$\sigma$ errors calculated using the method described in Section \ref{danny}.  The number of stars in each bin is shown in brackets in the legend.  Also shown are the $\chi^2$ and p-values for the comparison between the two samples in each panel.  We see a clear difference between the Splash and the high-$\upalpha$ disc across all metallicity bins.  Noteworthy is the difference in the $\upalpha$-abundances, the Splash having higher values for each $\upalpha$-abundance. This suggests that the Splash is an older population.  There is also a difference in Mn, an iron-peak element, with the Splash having lower [Mn/Fe] across each metallicity bin, suggesting again that it is older than the high-$\upalpha$ disc.}
    \label{chemcomp}
\end{figure*}

Fig.~\ref{chemcomp} shows the result of these comparisons,  each panel showing the median of each abundance for the two samples, and their respective errors. The medians indicate qualitatively the similarities across the various abundance ratios.  Quantitative estimates are obtained through computation of $\chi^2$ differences, which were computed as:

\begin{equation}
    \chi^2 = \sum_i\frac{\big({\rm\overline{[X/Fe]}}_{i,\rm{Splash}}-{\rm\overline{[X/Fe]}}_{i,\rm{Disc}})^2}{\sigma_{{\rm{[X/Fe]}}_{i,\rm{Splash}}}^2+\sigma_{{\rm{[X/Fe]}}_{i,\rm{Disc}}}^2},
\end{equation}

\noindent where ${\rm\overline{[X/Fe]}}$ is the median and $\sigma$ is the standard deviation of the medians for the 1,000 random samples.  The results are displayed in Fig.~\ref{chemcomp}, where both $\chi^2$ and the corresponding probability for the two distributions being the same are shown at the top of each panel.

\subsection{Second method: ${\bf \chi^2}$ distribution} 
\label{dom}

Another way to check whether there is a difference or similarity between our Splash and high-$\upalpha$ disc samples is by comparing the [X/Fe]--[Fe/H] relation in the two populations, for each element.  This method was developed by \cite{taylor2022}.  While in that paper the statistic adopted was the difference between the abundances in two populations, here we focus on the $\chi^2$.  This will give us an idea of which elements show the largest differences across the full metallicity range of our sample.  To calculate the $\chi^2$ for each abundance ratio, we split our samples into 10 bins so that each bin consists of about 60 Splash stars.  Each bin has a size that is at least twice the average measurement error of stars within.  The metallicity bins adopted are the following: [--1.1, --0.8], [--0.8, --0.71], [--0.71, -0.65], [--0.65, --0.62], [--0.62, --0.58], [--0.58, --0.54], [--0.54, --0.51], [--0.51, --0.47], [--0.47, --0.41], [--0.41, --0.3].  We then compute the $\chi^2$ using the following equation: 

\begin{equation}
\label{eqchi2}
\chi^2 = \sum_i\frac{(\overline{S_i}-\overline{D_i})^2}{\sigma_{S_i}^2+\sigma_{D_i}^2},
\end{equation}

\noindent where $\overline{S}$ and $\overline{D}$ are the median abundance ratios of the Splash and high-$\upalpha$ disc, respectively, per metallicity bin, $i$, and $\sigma$ is the standard error calculated as the standard deviation divided by the square root of the number of stars, $\frac{std}{\sqrt{N}}$.  
To aid in the interpretation of the $\chi^2$ values, we compare them with the null hypothesis estimated from the calculation of $\chi^2$ values obtained from the comparison of high-$\upalpha$ disc stars with randomly selected samples of the same population. In other words, we repeat the $\chi^2$ calculation, but now replace the Splash stars with random same-sized samples of high-$\upalpha$ disc members.  This is done $1,000$ times following the bootstrap method, which gives us a distribution of $\chi^2$ peaking at the value which is most similar to the whole high-$\upalpha$ disc sample.  This way, any $\chi^2$ value from the Splash and high-$\upalpha$ disc comparison that deviates from that of a random sampled distribution indicates a clear difference in the elemental abundance. This is the case across most of the metallicity range studied here.

\section{Results}
\label{sec:result}

In this section, we present results from the two methods used to compare the Splash with the high-$\upalpha$ disc.  Results from the first method are given in Section~\ref{dannyresult}, and those from the second method are in Section~\ref{domresult}.

\subsection{Abundance comparison results}
\label{dannyresult}

The results obtained from the first method, described in Section \ref{danny},  are displayed in Fig.~\ref{chemcomp}. The points represent the median values of the Splash (orange) and high-$\upalpha$ disc (black) for the respective abundance ratios within each metallicity bin, which is specified on the top of each panel.  We also show as the shaded region the respective $1~\sigma$ error obtained from $1,000$ bootstrapped samples.  
The $\chi^2$ value obtained from comparing the elements of the two populations per metallicity bin is also shown in each panel, along with the respective probability that the two populations are extracted from the same parent sample.  

We find that there is a significant difference between the two populations, across all metallicity bins. This is indicated by the large $\chi^2$ and zero probability.  This suggests that the Splash is a chemically distinct population from the high-$\upalpha$ disc.  We present more detailed comparisons for each element in the following sub-sections.

It is essential that we emphasise the high quality of APOGEE data, which makes possible the measurement of very small chemical composition differences between two distinct populations. 
This capability is due to the combination of the high quality of the APOGEE data on a star-by-star basis (high S/N, moderately high-resolution spectra, obtained with two twin, single-configuration, exceedingly stable spectrographs) with the targeting of a very large and homogeneous sample.  The latter is a very important aspect of the survey.  

After cleaning from contamination by foreground dwarfs (and to a much lesser degree, by AGB stars, wherever possible and necessary), the APOGEE main sample consists entirely of first-ascent red giants and red-clump stars.  Objects in these evolutionary stages cover a relatively narrow range of effective temperature, surface gravity, and micro-turbulent velocity.  
Such a state of affairs simplifies the spectral analysis tremendously, minimising the errors due to uncertainties in model atmospheres and line opacities, and ultimately leading to the generation of a massive database of stellar parameters and chemical compositions of unparalleled precision.

\subsubsection{$\alpha$-abundances}
\label{alpha}

First, we compare the $\upalpha$-abundances in the Splash and high-$\upalpha$ population. This includes the first five elements in Fig.~\ref{chemcomp}, as well as combining them all, i.e., the [$\upalpha$/Fe] ratio.  This abundance ratio effectively measures the relative contribution of SNe~II and SNe~Ia to the gas that stars formed from. This is because $\upalpha$-element enrichment originates from the explosions of massive ($\geq8~\rm{M}_\odot$) stars which undergo core collapse. Due to the very short lifetimes ($\tau_{\star}\simeq3-30~\rm{Myr}$) of these progenitors, the delay time distribution (DTD, distribution of times between star formation and the likely time of some event) of SNe~II for a burst of star formation is very narrow \citep[see figure~1 of][]{de_los_reyes_22}.  As a result, SN~II are thought to initially dominate chemical enrichment and thus being strongly tied to ongoing star formation.  Unlike SNe~II, type~Ia~SNe are not thought to occur so promptly following star formation.  This is because the delay time distribution of SNe Ia is set by the timescales of both stellar and binary evolution as their progenitors are intermediate-mass stars undergoing mass transfer. This results in a much broader range of delay times \citep{graur_14}, so that SNe~Ia are thought to dominate Fe-enrichment at later times, contributing minimal $\upalpha$-elements. 

Consequently, the relation between [$\upalpha$/Fe] and [Fe/H] is a useful diagnostic of a galaxy's history of star formation and chemical enrichment. It thus is expected that, for an extended history of star formation, it should be characterised by a flat sequence of [$\upalpha$/Fe] at low [Fe/H] connected to a declining sequence of [$\upalpha$/Fe] at high [Fe/H] by a so-called `$\upalpha$-knee'. 
The knee is thought to indicate the point in the chemical enrichment history when SNe Ia begin to contribute significantly to the chemical enrichment and thus its absolute value in [Fe/H] is thought to constrain the early star formation rate. The slope of a sequence in the $\upalpha$-Fe plane reflects the relative contributions to chemical enrichment of SNe~II and SNe~Ia.  A horizontal plateau reflects the IMF-averaged yield of $\upalpha$-elements and Fe contributed by SNe II {\it only}. Conversely, a negative slope of the shin indicates the contribution of SNe~Ia and SNe II. Outflows play a role in determining the slope of the shin for a given formation history. This is because if outflows are more efficient, by the time SNe Ia begin to contribute more gas which was predominantly polluted by SNe II will have been removed from the system (\citealp{tolstoy2009, andrews_17,mason2024}). 

For the full range of metallicities compared here, there are clear differences between the Splash and the high-$\upalpha$ populations.  The Splash has higher $\upalpha$-abundances than the high-$\upalpha$ disc, across all elements and all metallicity bins, with the exception of [O/Fe] in the lowest metallicity bin.  This indicates that the Splash is an older population, a result which is in agreement with other findings (\citealp{gallart2019, belokurov2020, grand2020, xiang2022}). This also supports the hypothesis that the Splash is part of the early heated disc.  
However, could there be an alternative mechanism, other than heating by a single accretion event, which could result in the higher $\upalpha$-abundance ratios of the Splash?  We explore this question in Section \ref{sec:grad}.

\subsubsection{Carbon and Nitrogen}
\label{lightoddz}
C and N abundances of stellar populations are useful diagnostics of the chemical evolution of a galaxy. The largest contributors of C and N to the ISM are massive stars which explode as SNe~II, followed by asymptotic giant branch (AGB) stars which form from intermediate-mass progenitors ($M\simeq1-11~{\rm{M}_\odot}$) \citep[AGB through SAGB;][]{renzini_81, siess_2010, karakas_2010}. As stars evolve along the RGB, the stellar envelope expands and it is shed \citep[see][and references therein]{wiersma_09a}. Prior to {being shed into the interstellar medium, the stellar envelope is enriched in the products of hot-bottom burning, by convective dredge-up \citep{renzini_81}}. Therefore abundances of C and N of a stellar population can reflect both contributions by SNe II {\textit{and}} AGBs with different lifetimes.

The differences in [C/Fe] and [N/Fe] are not as clear cut as what we see for the $\upalpha$-abundances.  For [C/Fe], there is a strong difference at the lowest metallicity bin, but the difference is not significant in the other bins. 
On the other hand, while [N/Fe] is significantly different in all metallicity bins, it changes from being higher in the Splash in the lower metallicity bins to being higher in the high-$\upalpha$ stars in the two most metal-rich bins.  By combining [C/Fe] and [N/Fe] we provide the comparison of [(C+N)/Fe] which minimises the effect of CNO mixing.  This abundance ratio does not exhibit a significant difference in any of the metallicity bins.
We also show the [C/N] ratio to provide an indication of the relative ages of the two populations, since it is observed that [C/N] correlates with age \citep[e.g.,][]{martig2016}.
However, the reliability of this correlation at low metallicities is poor.  In the higher metallicity bins, the Splash shows higher [C/N] ratios than the high-$\upalpha$ disc, which indicates that at these metallicities it is an older population.  To fully understand the result across all metallicity bins, chemical evolution models of [C/N] against [Fe/H] for two systems, one with an early quenched star formation and another one with an extended star formation, are required.

\subsubsection{Odd-Z elements}
\label{oddz}
Odd-Z elements, such as aluminium, are mostly synthesised by hydrostatic burning in massive stars and are released into the ISM by SNe~II. At low [Fe/H], the Milky Way's {\textit{in-situ}} populations and satellite galaxies show similar sub-solar odd-Z abundance ratios. At [Fe/H]$\gtrsim$--1, however, dwarf galaxies have characteristically depleted abundance ratios and the Milky Way generally shows super-solar abundance ratios \citep[see, e.g.,][and references therein]{tolstoy2009}. This can be explained by a combination of the metallicity-dependent yields of odd-Z elements with a higher and more extended history of star formation of the Milky Way than those of its satellites. Much work has been done to characterise the accreted substructures of the Milky Way using odd-Z abundances \citep[e.g.][]{hawkins_15,das_2020,horta2021,hasselquist_21,fernandes_23}.

The two odd-Z elements for which we make a comparison are aluminium and potassium.  Both [Al/Fe] and [K/Fe] show a very similar trend between the Splash and high-$\upalpha$ populations.  Other than the lower metallicity bin, where the two populations are the same for these abundance ratios, the Splash consistently has a higher abundance than the high-$\upalpha$ population and the difference increases at higher metallicities. The trend seen here indicates again that the Splash is an older population, due to the higher contribution from massive stars.  A high value for [Al/Fe] supports the notion that the Splash does not originate from a dwarf galaxy.

\subsubsection{Iron-peak}
\label{ironpeak}
The Fe-peak elements (e.g. Mn, Fe \& Co) are contributed to the ISM by both SNe~II and SNe~Ia \citep[e.g.][respectively]{portinari_98, iwamoto_99}. Some are also returned to the ISM by AGB winds in the late-stage evolution of intermediate-mass stars \citep{wiersma_09a}. As the relative contribution of SNe~II and SNe~Ia varies element-by-element, comparing the relative abundances of some Fe-peak elements can help constrain the importance of each channel as a function of metallicity, which places further constraints on the star formation history, in addition to those we discussed in Section~\ref{alpha}. A detailed discussion of the importance of these comparisons can be found in \cite{horta2023}.

For the iron-peak elements we mainly focus on Mn and Ni, but also show the results for Cr.  Throughout the whole metallicity range the [Mn/Fe] ratio is lower for the Splash, by a significant amount, which agrees with what we see in the alpha elements that the Splash is a less evolved population than the comparatively lower eccentricity high-$\upalpha$ disc.  For Ni, while at the lower metallicity, it is lower for the Splash, at higher metallicities no significant difference between the two populations exists. 

\subsubsection{Cerium}
\label{sprocess}

Neutron capture elements ($\rm{Z}\gtrsim30$) are formed by the capture of neutrons by the nuclei of the Fe-peak elements. Depending on the relative timescales of neutron capture and $\beta$-decay, such elements are termed slow or rapid ({\textit{s}}- or {\textit{r}}-) process elements. $s$-process elements originate from the atmospheres of intermediate-mass stars during the AGB phase of their stellar evolution, the most massive such progenitors have lifetimes on the order of 100 Myr.
 
From APOGEE we have only one {\textit{s}}-process element with reliable abundance ratios, Ce.  In our comparison, Ce does not show a clear trend between the Splash and high-$\upalpha$ population.  The Splash shows higher abundances ratios at the lower two metallicity bins, whereas the high-$\upalpha$ disc shows higher abundances ratios at the higher metallicity bins.  This is in agreement with previous results, showing that the star formation in the high-$\upalpha$ disc is more extended than in the Splash, as evidenced by the higher AGB contribution.

\begin{figure}
    \includegraphics[width=\columnwidth]{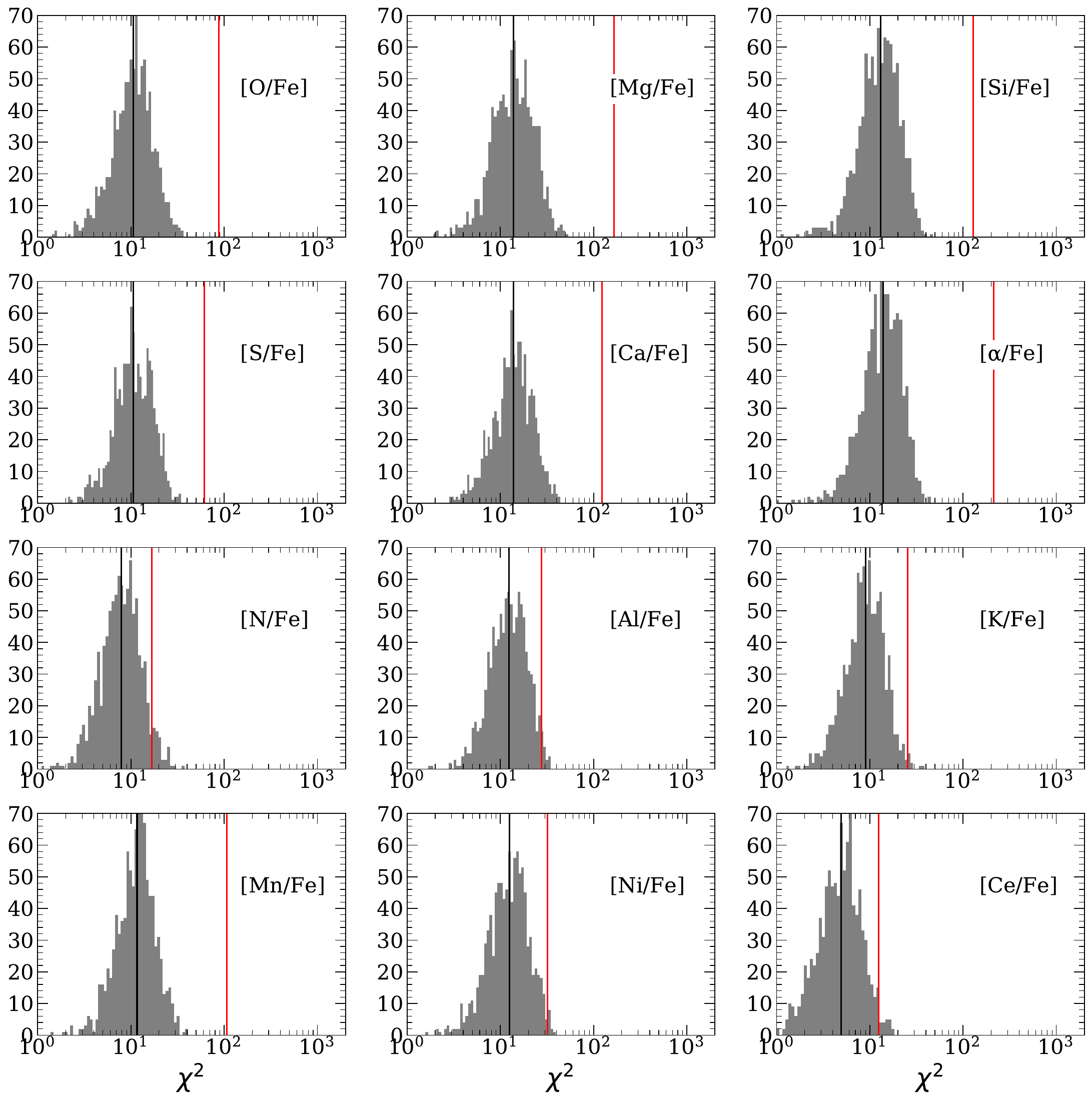}
    \caption{Comparison of $\chi^2$ between the Splash and high-$\upalpha$ disc (red lines) compared to the distribution of $\chi^2$ obtained from a random sample of the high-$\upalpha$ disc (grey histograms), with the medians of distributions shown as black lines.  These are computed in the range $-1.1<\rm{[Fe/H]}<-0.3$.  
    See Section \ref{dom} for a full description and Section \ref{domresult} for interpretation.  Shown here are only the abundances for which the differences in $\chi^2$ for the Splash compared to high-$\upalpha$ disc are large.
    All $\upalpha$-abundances show differences between the Splash and the disc, which confirms the result in the previous section.  We also see differences for N, Al, K, Mn, Ni and Ce.}
    \label{chi2}
\end{figure}
\begin{figure*}
    \centering
    \includegraphics[width=\textwidth]{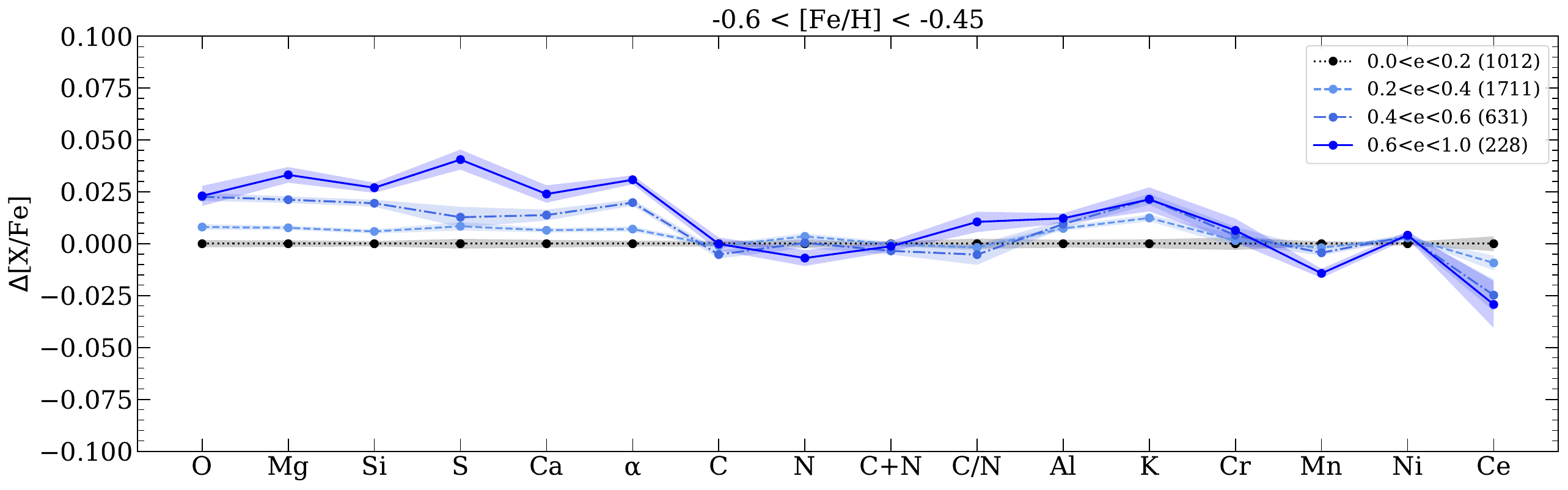}
    \caption{Chemical abundance comparison for  high-$\upalpha$ disc similar to Fig.~\ref{chemcomp}, but now the sample is split into four eccentricity bins, three bins of size 0.2, and one bin ($0.6<e<1$) corresponding to Splash. Only shown for the metallicity range with the largest sample.  The lowest eccentricity bin ($0.0<e<0.2$), shown in black, is subtracted from the others, including itself, which is why it is placed at a $\Delta$[X/Fe]$=0$. A clear trend is seen for some of the abundances with eccentricity. Most $\upalpha$ abundances increase with eccentricity, whereas Mn decreases. Though not shown, these trends are stronger at higher metallicities.  In most cases the Splash ($0.6<e<1.0$) has the most extreme abundances.}
    \label{echem_1}
\end{figure*}

\subsection{${\mathbf \chi^2}$ distribution results}
\label{domresult}
In this section, we show the results for the second method used to compare the Splash and high-$\upalpha$ disc, as described in Section \ref{dom}.
This statistic compares the relation between the abundance ratio of a given element, [X/Fe] and [Fe/H].  The statistic corresponding to the null hypothesis is estimated by running a comparison between 1,000 randomly selected disc samples with their parent population.  

Fig.~\ref{chi2} shows only those abundance ratios for which the difference in $\chi^2$ is large.  The $\chi^2$ value for the comparison between the Splash and the high-$\upalpha$ disc is shown as the red vertical line and the median of the distribution is shown as the black vertical line. 
The large difference between the chemistry of the Splash and high-$\upalpha$ is seen for 12 out of the 16 abundance ratios which we compare. 
There is a clear difference for all of the $\upalpha$ abundances shown, as well as a high difference in Mn, which highlights the distinction in the evolutionary stages of the two populations.  There are also smaller but significant differences in N, Al, K, Ni and Ce.

This method of checking the differences between our two populations supports the findings in Section \ref{dannyresult}, that the Splash has a statistically 'older' chemistry than the high-$\upalpha$ disc.
Our results highlight the level at which we can carry out statistical analysis to perform chemical tagging with such large data sets.  For example, in the [Al/Fe]--[Fe/H] plane for the Splash and high-$\upalpha$ (see Fig.~\ref{x-fe}) it is difficult to distinguish a large difference by eye.  However, with such a large high-precision data set, we can analyse them statistically and measure the difference.

\subsection{Is the Splash distinct from the high-$\upalpha$ disc?}
\label{sec:grad}
\begin{table*}
    \centering
    \resizebox{17.8cm}{!}{
    \begin{tabular}{|l|c|c|c|c|}
    \hline
    Comparison & {\bf $-1.1<{\rm [Fe/H]}<-0.8$} & {\bf $-0.8<{\rm [Fe/H]}<-0.6$} & {\bf $-0.6<{\rm [Fe/H]}<-0.45$} & {\bf $-0.45<{\rm [Fe/H]}<-0.3$}\\
    \hline
    $0.0<e<0.2$ vs. $0.2<e<0.4$ & 12.02 (0.44) & 46.37 & 105.77 & 52.4\\
    $0.2<e<0.4$ vs. $0.4<e<0.6$ & 22.1 (0.11) & 118.81 & 215.01 & 142.48\\
    $0.4<e<0.6$ vs. $0.6<e<1.0$ & 57.34 & 55.06 & 76.4 & 56.96\\
    \hline 
    \end{tabular}{}
    }
    \caption{$\chi^2$ comparison between each eccentricity bin for adjacent metallicity bins.  Non-zero p-values are shown in brackets.}
    \label{x2table}
\end{table*}

Results from previous sections suggest that the Splash is a distinct component from the rest of the disc in a wide range of chemical abundances, in addition to previously known differences in terms of Fe and $\upalpha$ abundances or kinematics. 
Although the evidence is strong, it is still possible that the chemical differences found here are the result of the specific criteria adopted to define the Splash.
Correlations between chemical composition and kinematics are known to exist (\citealp{mackereth2019, belokurov2020, horta2021}), so there is a possibility that the different chemical imprint of our Splash sample reflects the extension into a high eccentricity regime of the trends in chemical abundances  with kinematics already present in the high-$\upalpha$ disc. 
However, if the Splash is caused by the merger with a single event such as GE/S (rather than being the result of secular heating of the high$\upalpha$ disc), one could reasonably expect to find a clear signature in the relationship between chemistry and kinematics, in the form, for instance, of a discontinuity in such relations.

To test the hypothesis that the Splash is a smooth continuation of the high-$\upalpha$ disc into high eccentricities, we perform another chemical abundance comparison similar to that discussed in Section~\ref{danny}, but this time splitting the high-$\upalpha$ sample into four eccentricity bins.  
The results are shown in Fig.~\ref{echem_1}, the highest eccentricity bin ($0.6<e<1$) corresponding to the Splash population. A clear trend is seen for some of the abundances with eccentricity. Most $\upalpha$ abundances show a clear positive correlation with eccentricity,  whereas Mn behaves in the opposite way.  
In most cases, the Splash population ($0.6<e<1.0$) shows the most extreme abundance differences.
 
We note that Fig.~\ref{echem_1} shows results only for the metallicity bin with the largest number of stars ($-0.6<$ [Fe/H] $<-0.45$), but similar results are obtained for other bins (in fact, at higher metallicities, the trends in $\upalpha$ abundances are stronger). 
However, three elements, N, Cr and Ce, appear to make an exception.  While Cr and Ce do not show the same trend with respect to eccentricity in all metallicity bins, N shows a positive correlation with eccentricity at the lower two bins and a more negative correlation at the higher bins.
In Table~\ref{x2table} we include the $\chi^2$ values, and the associated non-zero p-values for the comparison between adjacent eccentricity bins.  
These show correlations between abundances and eccentricities for some elements, particularly the $\upalpha$-abundances and Mn.  

These results suggest that chemistry of the Splash population places it on the high end of the high-$\upalpha$ disc distribution of [$\upalpha$/Fe]. While the Splash appears to be chemically distinct from the whole disc, it is because it is more similar with a subset of the high-$\upalpha$ disc with high eccentricities. Differences between different eccentricity bins are also seen - this being made possible only by the high precision of APOGEE data.  If, as expected, stars with higher $\upalpha$-abundances are older, this indicates that there is also an age correlation with eccentricity, at least for the high-$\upalpha$ disc.  
This correlation was observed by \cite{belokurov2020}, in the distribution of $\rm{v_\phi}$ versus [Fe/H], colour-coded by median [$\upalpha$/Fe]  and by median age.  These authors found that both the median [$\upalpha$/Fe] and median age correlate with $\rm{v_\phi}$ in the thick disc region of the $\rm{v_\phi}$ -- [Fe/H] plane, the Splash being the low-angular-momentum part of the thick disc and therefore older than the disc.  We discuss the ages of Splash and high-$\upalpha$ disc further in Section~\ref{sec:mergers_Splash}, in the context of comparisons with simulations.

However, we caution that, for a full understanding of the chemical abundance trends, one also needs to take into account the effects of radial migration. \citet{ratcliffe2023} studied the effects of radial migration by deriving the birth radius of stars in the disc and showed, in agreement with our findings, that the oldest stars of the MW disc have higher $\upalpha$-abundances.  A study of radial migration is beyond the scope of this paper, and therefore we refer the reader to \citet{sharma2021} \& \citet{ratcliffe2023} for an insight into the time-dependent evolution of chemical abundances in relation to radial migration in the disc.

Our results show commonalities between the chemo-dynamical properties of the Splash and of the extreme end of the high-$\upalpha$ disc. Moreover, this type of investigation may yield information on the type of progenitor galaxy or galaxies responsible for the formation of the Splash. The general consensus is that the Splash is the result of a single catastrophic event, with the GE/S progenitor. In this case, one would expect an abrupt change in the correlation between chemistry and kinematics (e.g., eccentricity or angular momentum) when transitioning from the disc to the Splash. 
However, this would not necessarily be the case if the Splash arises from several minor mergers, which may dilute the signature of transition. To investigate the potential transition in chemo-dynamical properties, as well as the possibility of more minor mergers generating Splash-like features, we turn to cosmological simulations.

\section{Splash-like populations in simulations}
\label{artemis_sample}

In this Section we use the \texttt{ARTEMIS} cosmological simulations \citep{font2020}, a suite which contains 45 zoom-in hydrodynamical simulations of Milky Way-mass galaxies, run in a WMAP $\Lambda$CDM cosmological model with the \texttt{Gadget-3} code (last described in \citealt{springel2005}). The subgrid physics includes prescriptions for metal-dependent radiative cooling in the presence of a photo-ionizing UV background, star formation, stellar and chemical evolution, formation of supermassive black holes, stellar feedback from supernova (SN) and stellar winds, and from active galactic nuclei. The chemical enrichment model tracks 11 element species (H, He, C, N, O, Ne, Mg, Si and Fe). 

The dark matter particle mass is $1.17 \times 10^5 {\rm M}_{\odot}h^{-1}$, while the initial baryon particle mass is $2.23 \times 10^{4} {\rm M}_{\odot} h^{-1}$. The total galaxy masses of the Milky Way-like haloes range between $0.8 < \rm M_{200} /{10^{12} \, {\rm M}_{\odot}}< 1.2$, where $\rm M_{200}$ is the mass enclosed within a volume of mean density 200 times the critical density at redshift $z = 0$. More details about the simulations can be found in \citet{font2020} (see also \citealt{font2021,font2022} for more detailed comparisons with observations of MW analogues). Merger trees for MW-mass hosts are generated using the \texttt{D-haloes} code, which is based on algorithms of \citet{srisawat2013} and \citet{jiang2014}, to track the progenitors and descendents of each subhalo in the simulations. Gravitationally bound structures (haloes and subhaloes) are identified using the \texttt{SUBFIND} halo finding algorithm (last described in \citealt{springel2001}).

In the following, we make a distinction between two categories of MW-mass systems, based on the ratio between the mass of MMAP and that of its host galaxy. Specifically, one category comprises galaxies that experience a GE/S-like event at early times, denoted as `MW-GES'. 
These are systems where the mass ratio of MMAP/host is $>0.4$ and accretion of the MMAP occurred $>8$~Gyr ago, as indicated in figure 7 of \citet{dillamore2022}. 
Four galaxies fall into this category: G29, G30, G34 and G42. The second category consists of galaxies whose mass ratio of MMAP/host is $<0.4$, but their MMAP still occurs more than $\sim 8$~ Gyr ago, i.e., around the same time as the GE/S in the MW. 
These are called `MW-MA', which stands for `minor (early) accretion'. The MMAP/host mass ratios in the MW-MA systems are $\sim 0.1 - 0.2$ (see figure 7 of \citealt{dillamore2022}). We note, however, that all systems selected here have high total accreted fractions (specifically, the contributions to the accreted population of the more radially anisotropic fitted components by \citet{dillamore2022} are $>40\%$). There are three galaxies in the MW-MA category: G17, G19 and G44. Both MW-GES and MW-MA galaxies display a disc morphology at present time. 

 With this subsample, we aim to check whether Splash populations exist in both or only in one of these categories (e.g., only in MW-GES), and to quantify the Splash fractions in each case.  For more direct comparisons with observations, we select star particles in the 'solar neighbourhoods' in these simulated discs, which are defined as before as regions with $|Z|<3$~kpc and $5< R <11$~kpc, in galactocentric cylindrical coordinates.  We also limit the metallicity to [Fe/H]$>-2.5$, to best match the properties of the observed sample.

\subsection{Relationship between chemistry and kinematics}
\label{comparison_Artemis1}

\begin{figure}
    \includegraphics[width=\columnwidth]{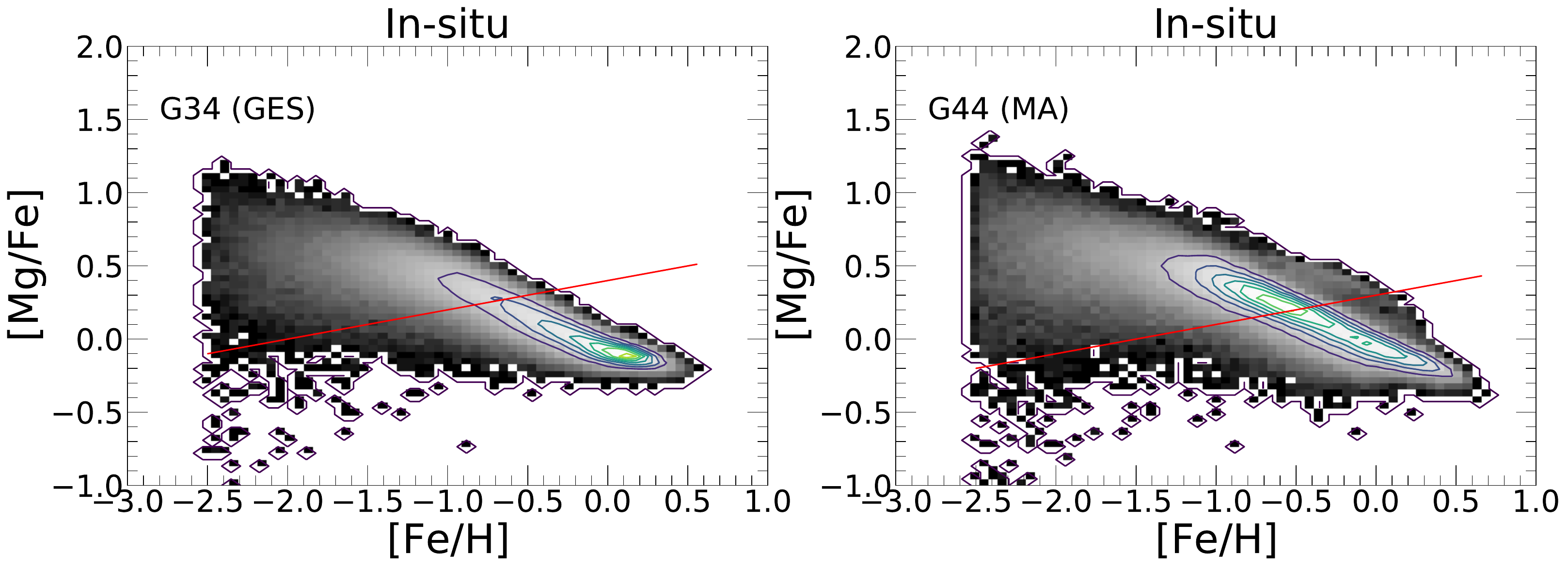}
    \caption{Examples of our selection of the high-$\upalpha$ disc in the [Mg/Fe]--[Fe/H] plane for a simulated MW-GES galaxy (G34, {\it left panel}) and a MW-MA galaxy (G44, {\it right panel}).  Data points above and to the left of the red line are referred to as high-$\upalpha$.  The selection is made only for the {\it in situ} populations which is shown as 2D histograms.  Also shown are the contours for the {\it in situ} populations, to highlight the dividing line between the high- and low-$\upalpha$ stars.}
    \label{mg-fe_artemis}
\end{figure}

We first inspect the relationship between the chemistry, specifically the [Mg/Fe] ratios, and kinematics in simulated galaxies. The choice of [Mg/Fe] is motivated by the clear correlation found earlier between the $\upalpha$-abundances and eccentricity in the high-$\upalpha$ MW disc.  An investigation of the simulations has the added bonus of being able to establish unequivocally which stars are accreted or {\it in situ}. In \texttt{ARTEMIS}, star particles are labeled as {\it in situ} if they formed within the halo of the MW-mass host, and accreted if they formed inside satellite galaxies {\it prior} to accretion (see \citealt{font2020} for details).

As a measure of kinematics, we use the angular momentum (Lz) instead of the eccentricity, which was adopted in the analysis of APOGEE data in previous sections. 
However, Lz correlates very well with eccentricity\footnote{We also verified that Lz correlates with orbital eccentricities in the simulations, and this turned out to be the case, as expected.} 
To select the high-$\upalpha$ disc in the simulated galaxies, we make a simple cut to the {\it in situ} population in the [Mg/Fe]--[Fe/H] plane of each galaxy that best reproduces the separation between the high- and low-$\upalpha$ discs in APOGEE. 
We thus remove the high-density region at low [Mg/Fe] and high [Fe/H] associated with the low-$\upalpha$ disc.  Fig.~\ref{mg-fe_artemis} shows this selection for two simulated galaxies, G34 (a MW-GES galaxy) and G44 (a MW-MA galaxy). 

\begin{figure*}
    \begin{center}
        \includegraphics[scale=0.24]{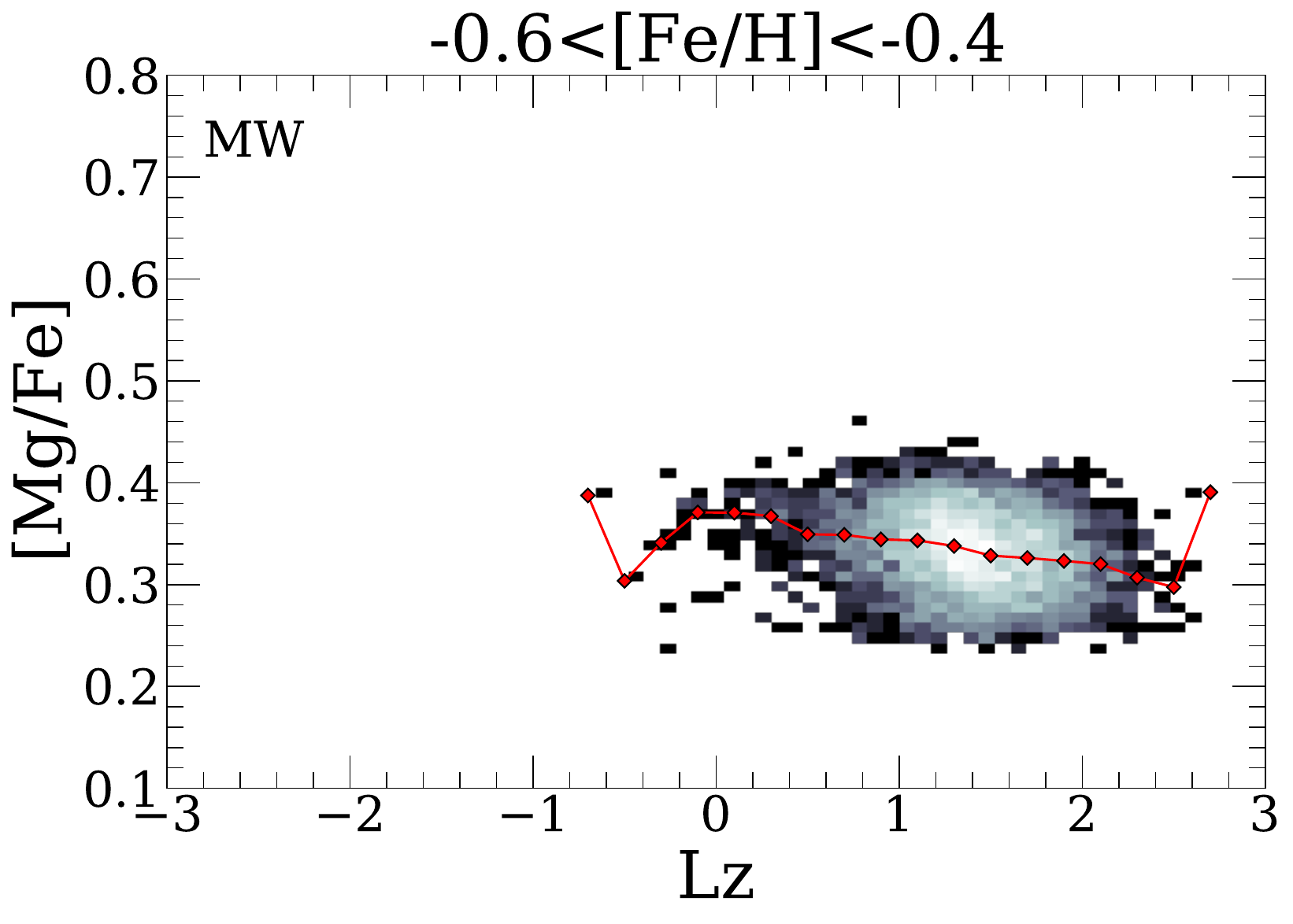}
    \end{center}
    \includegraphics[scale=0.22]{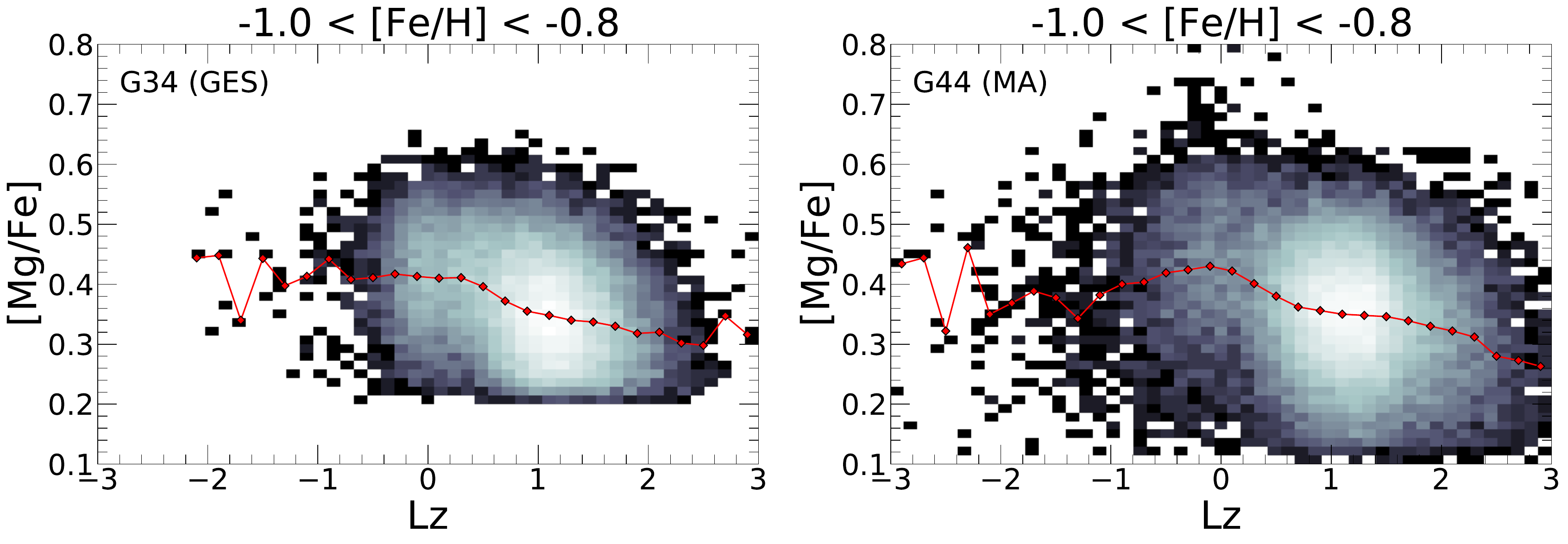}
    \caption{{\it Top panel:}  [Mg/Fe]--Lz distribution for the MW's high-$\upalpha$ disc in the range of $-0.6<$[Fe/H]$<-0.4$.  {\it Bottom panels:} The two simulated galaxies shown in Fig.~\ref{mg-fe_artemis}, G34 (MW-GES) and G44 (MW-MA), this time shown in the [Mg/Fe]--Lz plane. The high-$\upalpha$ population is shown as 2D-histograms. [Fe/H] is restricted to a narrow range, to avoid any evolutionary biases.  The running medians, shown with red lines, suggest that the correlations between [Mg/Fe] and Lz are similar between the MW-GES and MW-MA galaxies, and with the MW.}
    \label{mg-lz_artemis}
\end{figure*}

\begin{figure*}
\begin{tabular}{cc}
    \includegraphics[width=0.9\columnwidth]{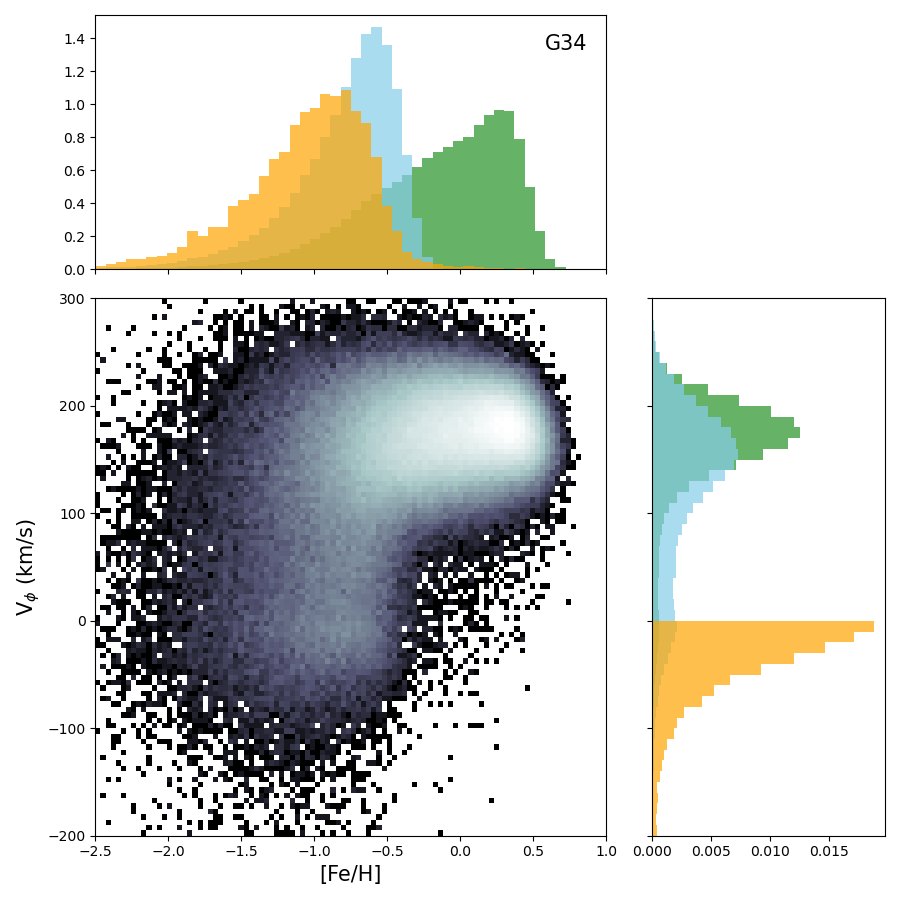} &
    \includegraphics[width=0.9\columnwidth]{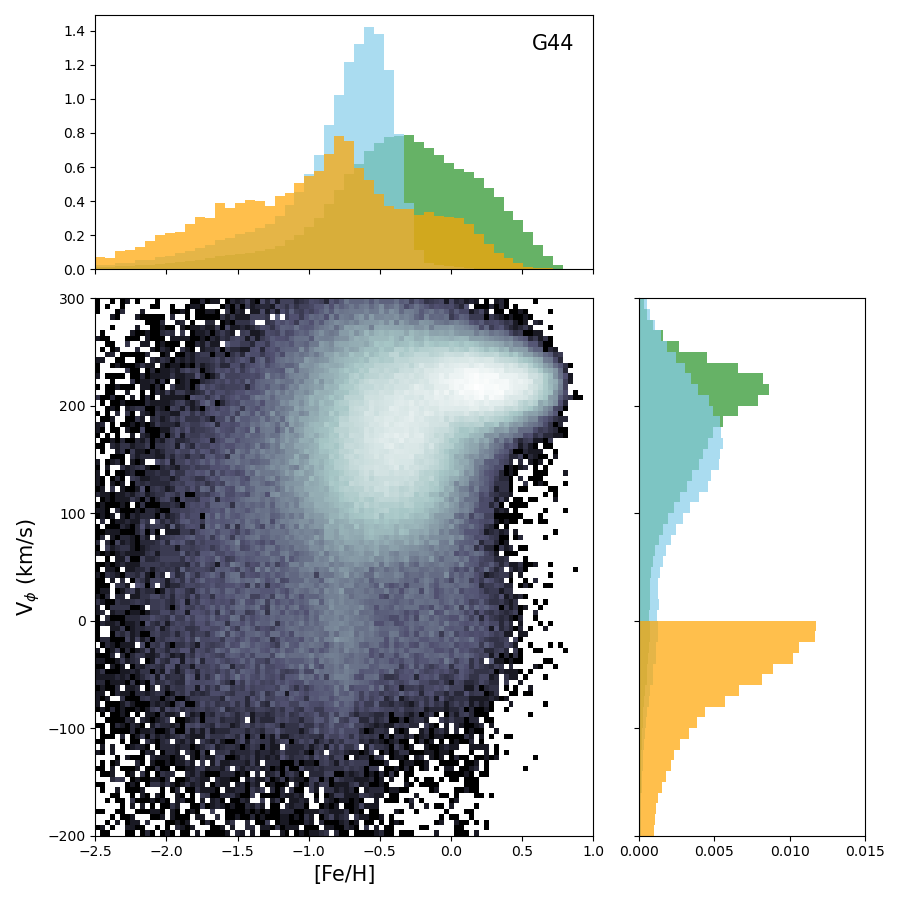}\\
\end{tabular}
    \caption{$v_{\phi} -$ [Fe/H] distributions in the G34 and G44 solar neighbourhoods ({\it main panels, on the left and right, respectively)}. Side sub-panels show the normalized distributions of the Splash (orange), high-$\upalpha$ disc (blue) and all disc stars (green). The Splash components are more metal-poor and more $\upalpha$-enhanced than the corresponding high-$\upalpha$ discs.}
\label{fig:FeH-vtheta}    
\end{figure*}

\begin{figure*}
\begin{tabular}{cc}
    \includegraphics[width=0.9\columnwidth]{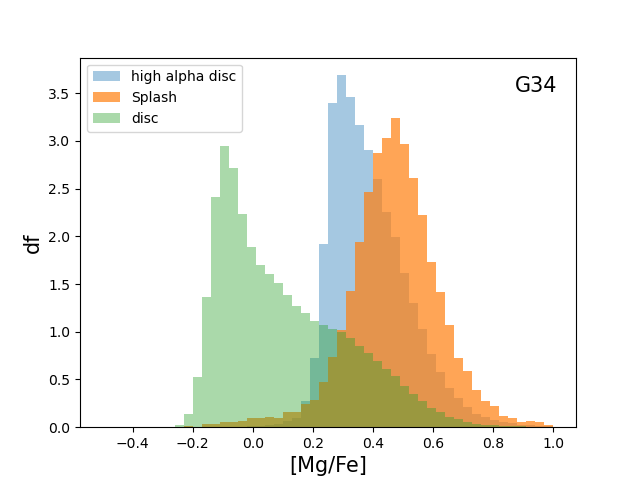} &
    \includegraphics[width=0.9\columnwidth]{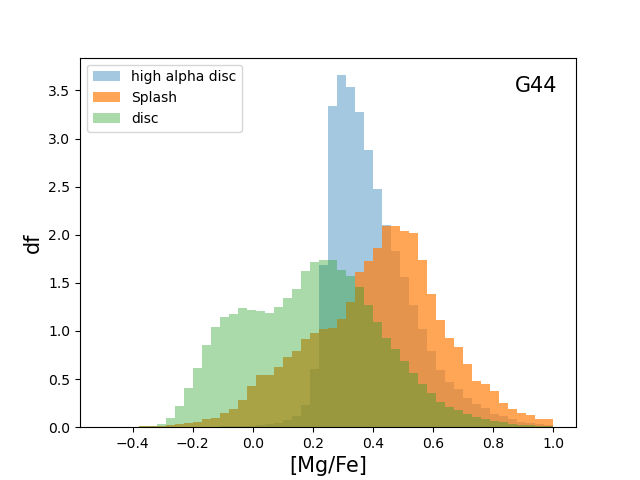}\\
\end{tabular}
\caption{Normalised [Mg/Fe] distributions for the Splash, high-$\upalpha$ disc and all disc stars in the solar neighborhoods of G34 and G44.}
\label{fig:MgFe}    
\end{figure*}
Fig.~\ref{mg-lz_artemis} shows the [Mg/Fe]--Lz distribution for the high-$\upalpha$ populations in these two galaxies and, for comparison, the same distribution in the MW.  All three galaxies show an anti-correlation between [Mg/Fe] and Lz. This is indicated more clearly by the running medians, shown with red lines in this figure. The slopes of the [Mg/Fe] -- Lz relations  are similar for both MW-GES and MW-MA galaxies, and also similar to the one in the MW itself. This suggests that the overall relation between chemistry (specifically, the $\upalpha$-abundances) and kinematics is not sufficient to distinguish the mass of the MMAP in a given host galaxy. In our sample of six simulated galaxies, five have similar [Mg/Fe] -- Lz trends (the exception is G30, which is a MW-GES galaxy).  

Interestingly, several of the simulated galaxies also display a discontinuity in the relation between [Mg/Fe] and kinematics (Lz), right at the transition between retrograde to prograde motions (see, for example, Fig.~\ref{mg-lz_artemis}). This discontinuity may be associated with a burst of star formation around the time of the disc spin-up (see also \citealt{belokurov2022,dillamore2023b}). However, we note that MW does not display such a discontinuity in the $\upalpha$-abundances, so further study is needed to understand this feature.

Fig.~\ref{fig:FeH-vtheta} shows the $v_{\phi} -$ [Fe/H] distribution of solar neighbourhood disc stars in the G34 and G44 (main panels), with the normalized distributions of stars in the Splash (orange), high-$\upalpha$ disc (blue) and all disc stars (green) on the side sub-panels.  Fig.~\ref{fig:MgFe} shows the corresponding [Mg/Fe] distributions for the three simulated components in these two galaxies.
The Splash components are clearly visible in both cases (MW-GES vs. MW-MA), and are distinct in terms of kinematics and chemical compositions from the high-$\upalpha$ discs. 
Specifically, the Splash stars are more metal-poor and more $\upalpha$-enhanced than the high-$\upalpha$ disc stars, in agreement with our findings for the MW data, discussed in Section~\ref{sec:result}. In addition, as shown below (Fig.~\ref{fig:ages}), the Splash components are also predicted to be typically older than the high-$\upalpha$ discs.

\subsection{Correlations between the Splash and the retrograde accreted fractions}
\label{comparison_Artemis2}

Under the assumption that the properties of the Splash depend on those of the MMAP (the putative cause of the Splash), several correlations have been investigated. 
For example, \citet{grand2020} show a correlation between the Splash fractions and the masses of the (GE/S) MMAP (see their figure 10), although the correlation is rather weak. 
Similarly, \citet{dillamore2022} find a strong correlation (Pearson coefficient of 0.8) between the mass fraction of the {\it in situ} retrograde stars (i.e., the Splash) versus the fraction of accreted stars within $5<r<30$~kpc (see their figure 16).

However, the total accreted component may include a mixture of prograde and retrograde stars, resulting from multiple mergers. These could be major mergers (as in the case of GE/S), or more minor ones. Here we seek a clearer connection to the progenitor(s) that may have caused the Splash. 
We therefore plot the Splash fraction against the fraction of {\it retrograde} accreted stars. To compute the Splash fractions, we select all {\it in situ} star particles on retrograde orbits (Lz$<0$) at redshift 0 and then compute the ratio between the {\it in situ} retrograde and prograde populations.  
In addition, we study the spatial variations of the Splash fractions by splitting the simulated `solar neighbourhood' annuli into six sections subtending the same azimuthal angle from the galactic centre. We then calculate the Splash fractions for each of them.  

\begin{figure}
    \centering
    \includegraphics[scale=0.3]{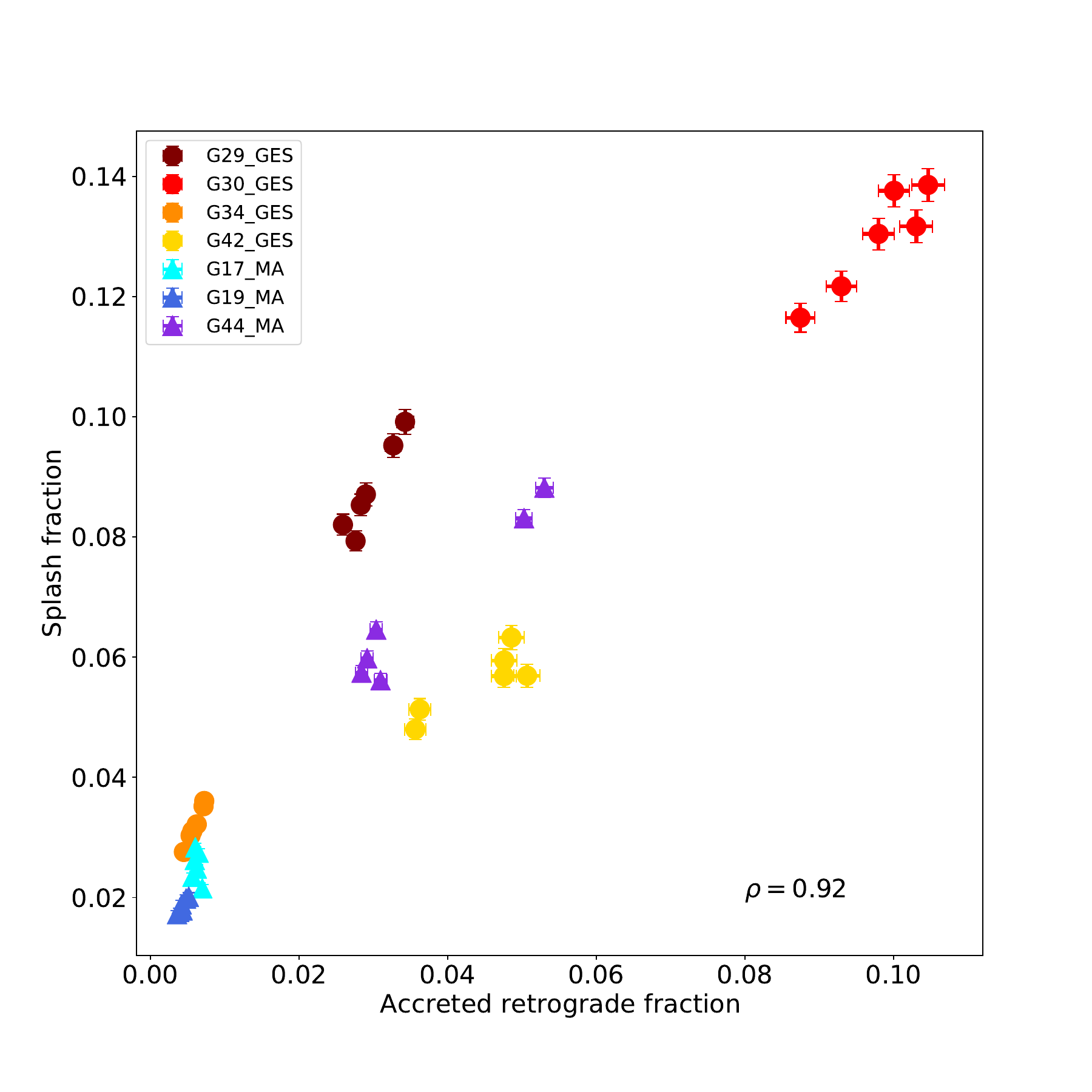}
    \caption{The relation between the Splash fractions and the accreted retrograde fractions for four MW-GES galaxies (circles) and three MW-MA galaxies (triangles), each galaxy having 6 measurements in equally spaced azimuthal sections.  Each colour represents a galaxy with the associated error bars. The Splash fractions show a strong positive correlation with the accreted retrograde fractions, confirmed by the Pearson correlation coefficient of 0.92, shown in the bottom right of the plot}.
    \label{splashfracvaccretrofrac}
\end{figure}

Fig.~\ref{splashfracvaccretrofrac} shows the results for the four MW-GES and three MW-MA galaxies, each with six measurements. This shows a strong linear correlation (Pearson coefficient of 0.92) between the Splash fractions and the fractions of accreted retrograde stars.  
This result should be contrasted with that by \citet{dillamore2022}, who showed that there is a good correlation with a lower Pearson coefficient, with the total accreted fraction.
The stronger correlation with retrograde accreted mass suggests that accretions with that orientation can contribute more strongly to the formation of a Splash population.
We also note that the fraction of retrograde accreted stars is computed in our study within the same solar neighborhood region as the Splash fraction, whereas \citet{dillamore2022} compute it globally in the host galaxy.

Fig.~\ref{splashfracvaccretrofrac} shows that, on average, MW-GES galaxies have higher Splash fractions than MW-MAs. They also have, on average, higher accreted retrograde fractions. However, this is not always the case: one MW-MA galaxy, G44, also has a high Splash fraction, whereas G34, a MW-GES galaxy, has a low value. This is because G44 has more mergers on retrograde orbits which increase the Splash fraction, despite their lower masses. Conversely, G34 has its MMAP on a prograde orbit and, while it still heats up the disc considerably, it does not produce a significant Splash, at least under the assumed definition for it.
Therefore, our results suggest that Splash-like features can be produced in a variety of ways: not only via GE/S-like mergers on retrograde orbits, but also by repeated minor mergers on retrograde orbits. We caution, however, that the orientation of the disc could change over time, as the disc tends to align with the orbital plane of a massive merger (see \citealt{dillamore2022});  this may affect the classification of retrograde stars, particularly in cases where the discs flips its orientation. We expect these cases to be rare though.

We note also that some (MW-GES) galaxies may have their Splash features located outside of the regions studied here (for example, concentrated within $R<5$~kpc or beyond $|Z|>3$~kpc). This also raises the possibility that a significant population of stars from the Splash may lay undiscovered yet in the MW. We discuss the spatial extent of the simulated Splash features in Section \ref{sec:heatsim}.

Furthermore, by inspecting all azimuthal regions individually, it is clear that the Splash fractions vary significantly from section to section. This suggests that the impact of the MMAP is often local in the disc. Interestingly, the two sections of G44 that have high fractions are situated opposite each other.  This suggests that measurements across various radial directions in the disc could provide information about the shape of the Splash.  Note also the grouping of the three galaxies around $(0.04, 0.07)$, which indicates the magnitude of the scatter in this relation.  We discuss the possible causes of this scatter in the next section.

\subsection{Spatial distribution of Splash features}
\label{sec:heatsim}

\begin{figure*}
    \begin{tabular}{cc}
    \includegraphics[width=0.4\linewidth]{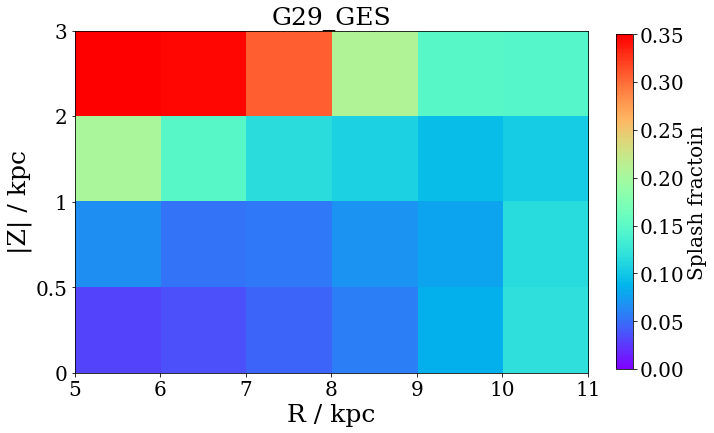} &
    \includegraphics[width=0.4\linewidth]{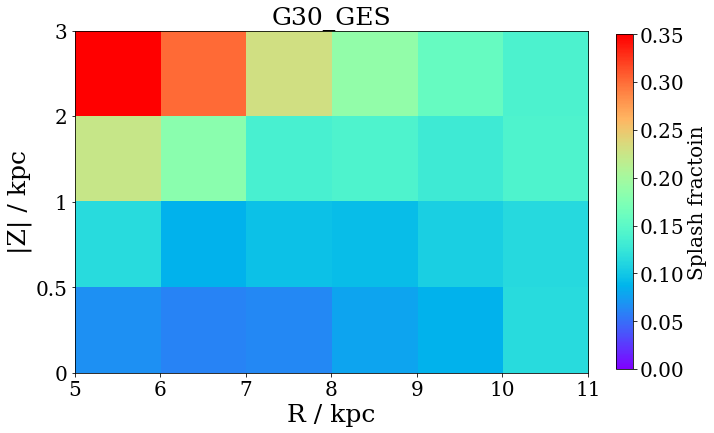} \\
    \includegraphics[width=0.4\linewidth]{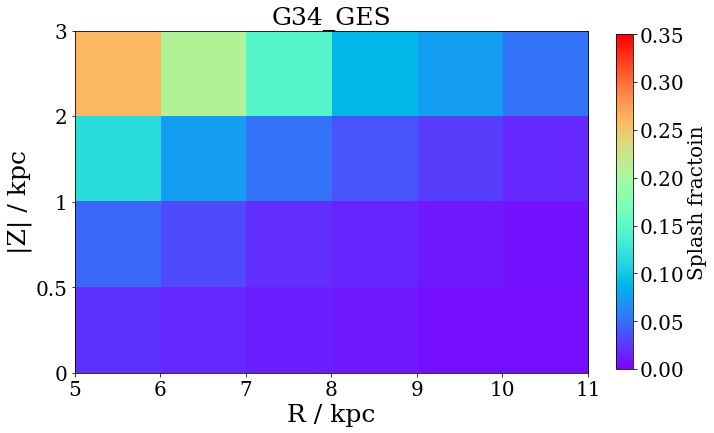}  &
    \includegraphics[width=0.4\linewidth]{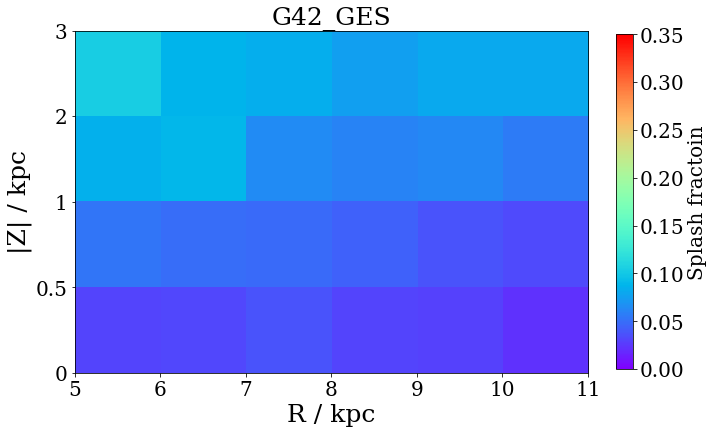} \\
    \includegraphics[width=0.4\linewidth]{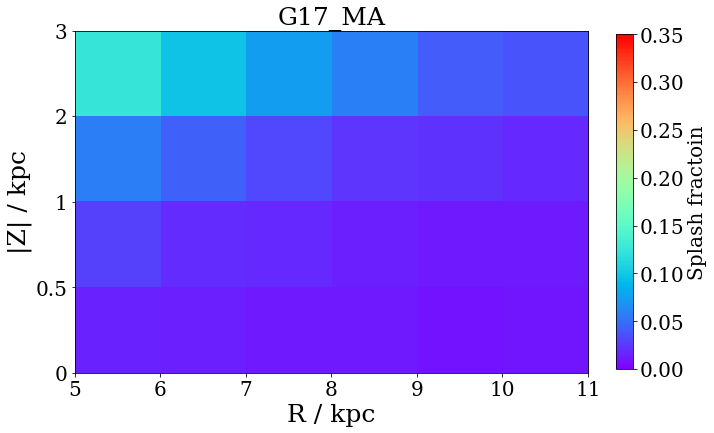} &
    \includegraphics[width=0.4\linewidth]{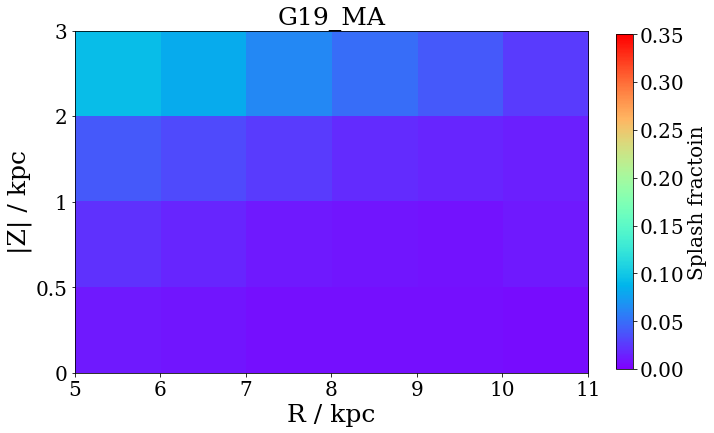} \\
    \includegraphics[width=0.4\linewidth]{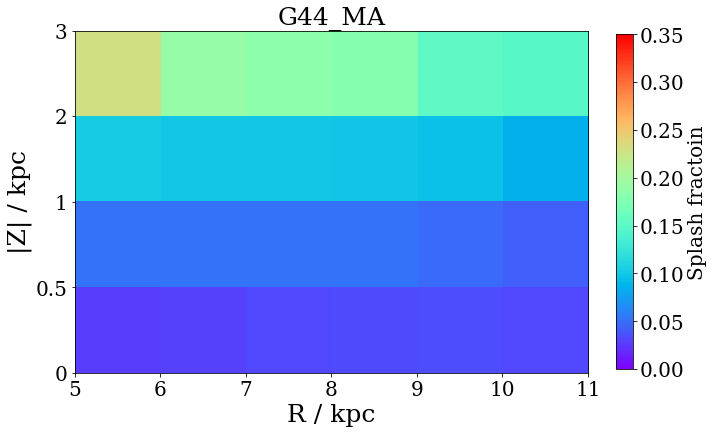} 
    \end{tabular}
    \caption{Confusion maps for the four of the MW-GES galaxies (G29, G30, G34 and G42) for three MW-MA ones (G17, G19 and G44). Splash fractions are computed in the solar neighbourhood regions, in a similar way as it was done for the MW (see Fig.~\ref{confusion}).}
    \label{ConfusionSims}
\end{figure*}

To illustrate the extent of Splash-like features in simulations, we also construct confusion maps as it was done for the MW (see Fig.~\ref{confusion}). These are shown in Fig.~\ref{ConfusionSims}. Generally speaking, we find that the Splash fractions follow the same trends across all galaxies, with higher values at lower $R$ and/or higher $Z$. Notably, this behaviour is present regardless of whether galaxies have had a GE/S-like event or not. 

There are, however, distinct differences between the confusion maps of MW-GES and MW-MA galaxies, in that the former have, on average, higher Splash fractions. 
The extent of the Splash population in MW-GES galaxies is also wider across the solar neighborhood regions than in the MW-MA ones, both radially and in height. 
The spatial distribution of Splash fractions in the simulated MW-GES galaxies is also in good qualitative agreement with the one measured in the MW (compare Fig.~\ref{ConfusionSims} with Fig.~\ref{confusion}). On an case-by-case basis, however, some MW-GES simulated galaxies display more extended regions (both in $R$ and $|Z|$) and with higher Splash fractions (e.g., G29, G30), while in others, the Splash population is confined to smaller regions and the Splash fractions are lower (e.g., G42). 
These variations may be explained by the difference in the merger histories (e.g., the masses of progenitors, the times of accretion, or their orbits).

\subsection{Merger histories and the onset of the Splash}
\label{sec:mergers_Splash}

In this section, we investigate in more detail the evolution of the Splash components in relation to past mergers and seek to identify the specific merger(s) responsible for their onset. We also investigate the possible starbursts associated with these mergers in the composition of the Splash and the disc.

\begin{figure*}
    \begin{tabular}{cc}
    \includegraphics[width=0.45\linewidth]{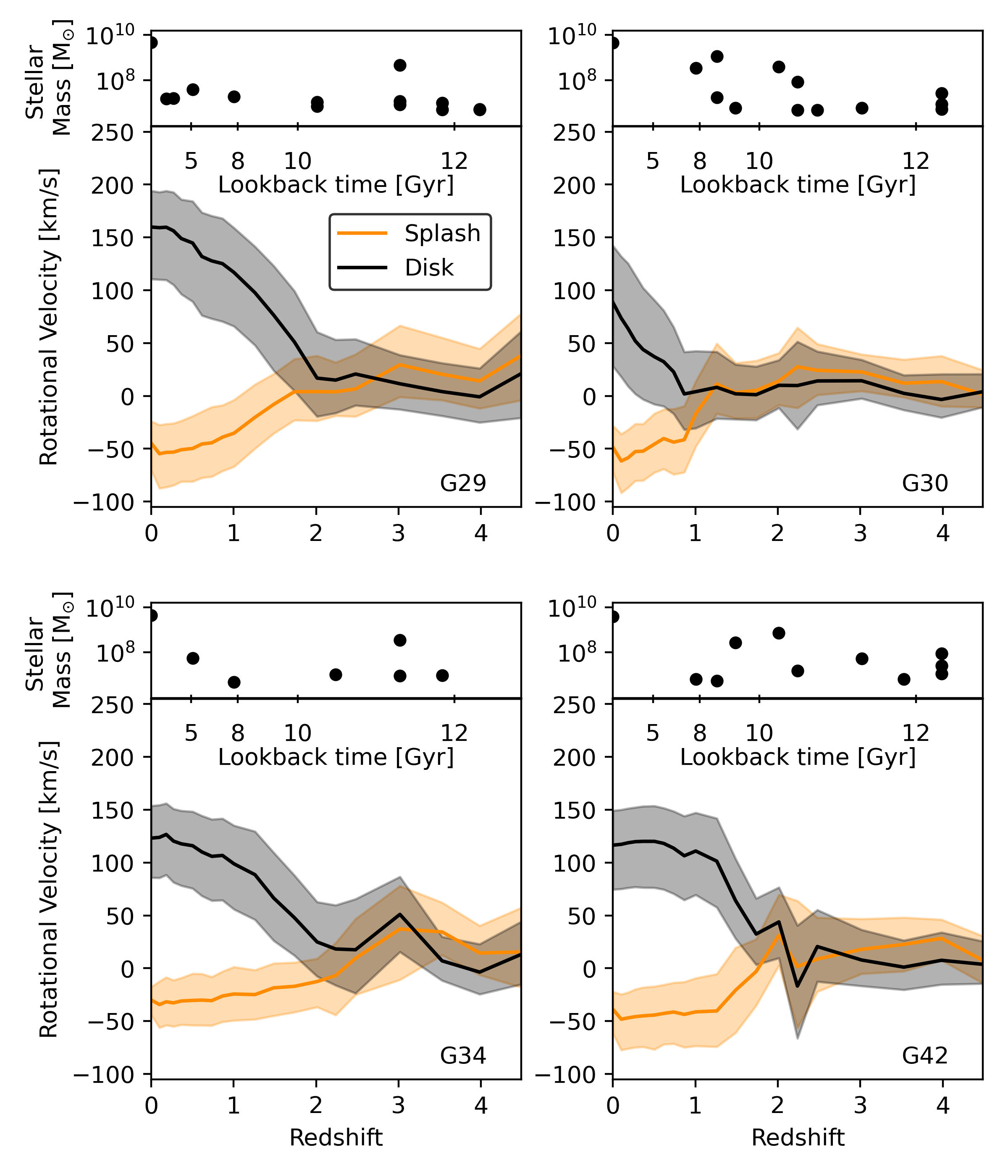} &
    \includegraphics[width=0.45\linewidth]{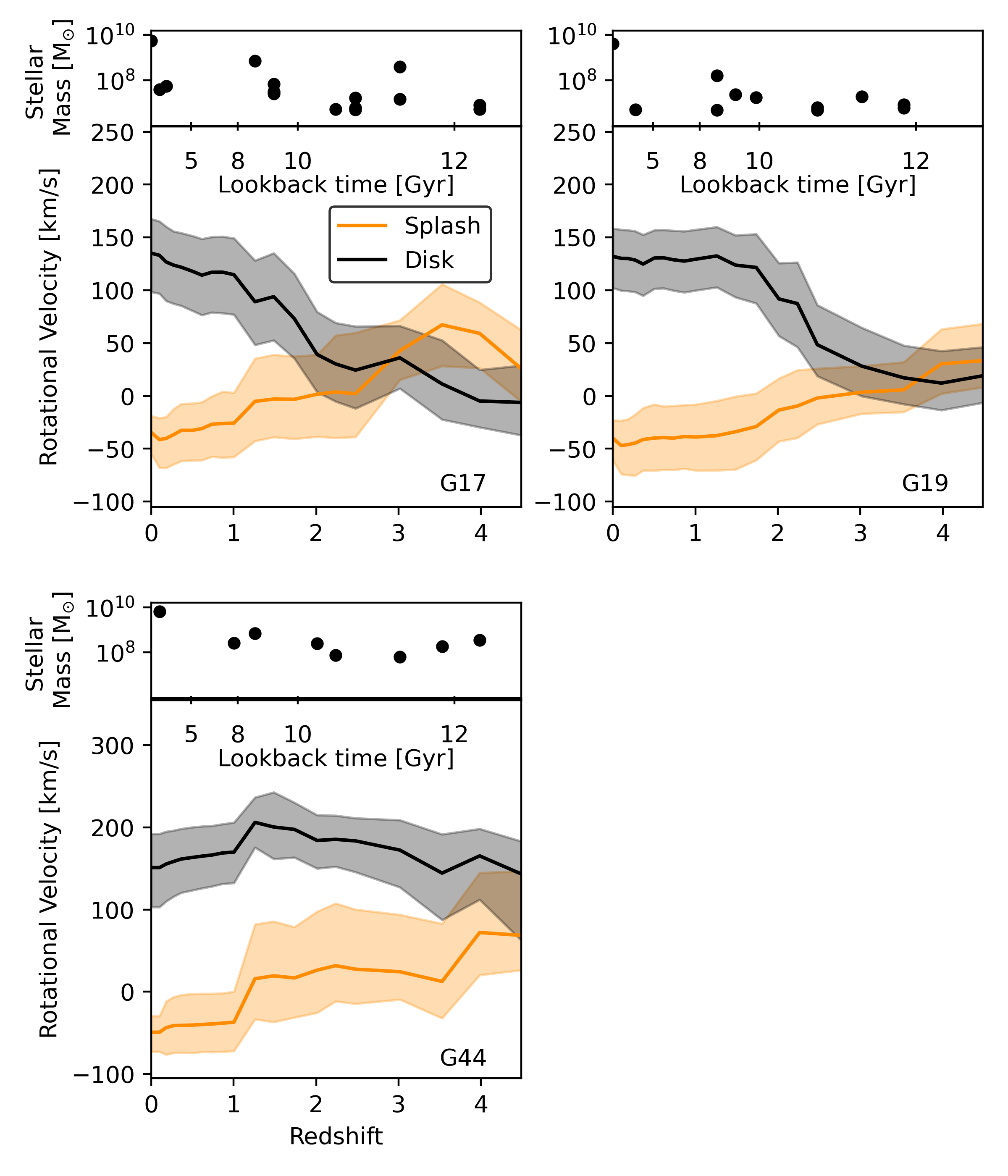}
    \end{tabular}
    \caption{Evolution versus lookback time (and redshift) of the rotational velocity ($v_{\phi}$) of the Splash (orange) and other disc stars (grey) in the solar neighborhood today. The top subpanels indicate the accretion history of each galaxy, with black circles corresponding to lookback time / redshift of accretion of progenitors and their stellar masses (measured at accretion).  The four panels on the left show the MW-GES galaxies, and the three on the right, the MW-MAs. The Splash population originates in the disc, and its onset is associated with a massive (stellar mass $>10^8\, {\rm M}_{\odot}$) merger. However, this merger is not always the MMAP, as seen from the Splash turnover towards retrograde orbits not always being coincident with the merger of the MMAP.}
    \label{fig:mergers}
\end{figure*}

Fig.~\ref{fig:mergers} shows the evolution of the rotational velocity $v_{\phi}$ for the Splash and disc stars in the simulated solar neighbourhoods in our sample. For this, we select all star particles in the two components at $z=0$ and trace them back in time in the simulations. 
At every snapshot, we re-orient the particles in the coordinate frame of the disc at that time, so that rotational velocities are always in the plane of the disc. The top sub-panels show the corresponding merger histories for these galaxies, expressed as stellar masses of progenitors (measured at the time of accretion onto their hosts) versus the accretion lookback-time (and also versus redshift). 
Galaxies have multiple mergers of stellar masses $>10^8 \, {\rm M}_{\odot}$, some on prograde and some on retrograde orbits relative to the axis of rotation of disc stars (at that time). Note though that not all mergers may leave an imprint in the solar neighborhood analysed here, as their impacts could be quite localised. Moreover, prograde mergers can still heat up the disc significantly, but their effects are not fully captured in the Splash population, as defined in this analysis.

This figure clearly shows that the Splash populations originate in the disc, as expected. However, the turnover towards retrograde orbits is not always coincident with the accretion of the MMAP (see also \citealt{dillamore2022}, e.g., their figure 15). 
This could be because the MMAP is on a prograde orbit, or even if it is on a retrograde one, its associated Splash may not be near the solar neighborhood. Nevertheless, we see that Splash-like features are ubiquitous in systems with mergers above $\sim 10^{8}\, {\rm M}_{\odot}$ in stellar mass.

We can now put more context into the unusual difference between G34 and G44 seen in Fig.~\ref{splashfracvaccretrofrac} (the former, a MW-GES galaxy with low Splash fraction, and the latter, a MW-MA with a higher Splash fraction). G34's MMAP, with a mass on the order of $10^9\rm{M_\odot}$, occurs at around a redshift of 3, however, the number of other massive accretions is relatively small. 
On the other hand, while the progenitor that generates the Splash in G44 is of lower mass and occurs early on (at redshift $\sim 4$), there are four more massive mergers (with masses of $\sim 10^9\rm{M_\odot}$) since redshift of 2, some of which may contribute to the final Splash fraction.

\begin{figure*}
\begin{tabular}{cc}
    \includegraphics[width=\columnwidth]
    {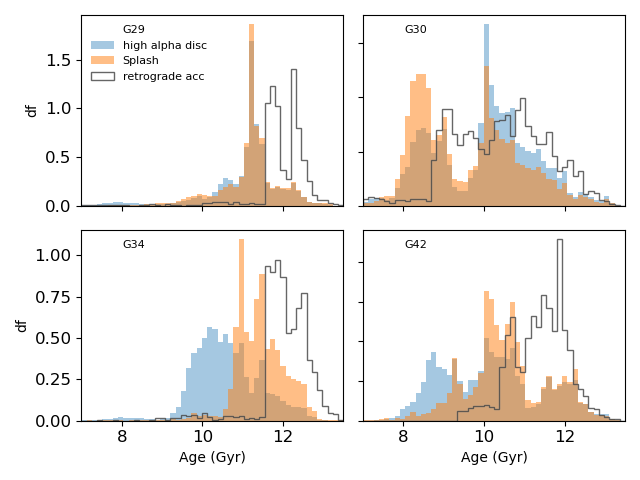} &
    \includegraphics[width=\columnwidth] 
    {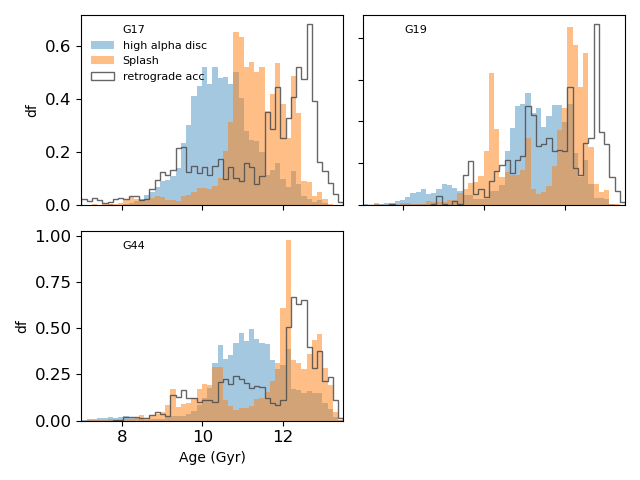}
    \end{tabular}
    \caption{The distribution of ages in the Splash (orange), high-$\upalpha$ disc (blue) and the retrograde accreted fraction (black histograms) in the simulated solar neighbourhoods. As before, MW-GES are shown on the left, and MW-MA on the right. In most cases, a clear association can be seen between the sharp decrease from the peak distribution in the retrograde stars (likely corresponding to the end of star formation in the progenitor, and the onset of the Splash). Also, Splash stars are typically older than the high-$\upalpha$ stars, indicating that they originate from the oldest population of the high-$\upalpha$ disc.}
    \label{fig:ages}
\end{figure*}

Fig.~\ref{fig:ages} shows the distribution in the ages of stars in the Splash components (in orange), contrasted with the ages of retrograde accreted stars (empty black histograms) and of the high-$\upalpha$ disc (blue), respectively. 
Focusing first on Splash versus retrograde accreted stars, we see that, in general, there is a sharp decrease in the distribution of retrograde accreted stars - which likely signals the end of star formation in the accreted progenitor due to ram-pressure stripping - followed shortly by a peak in the Splash distribution. 
This adds more evidence to the association of the two events. In some cases  (e.g., G30, G44), the distribution of ages in the retrograde population is spread more widely, and this appears to be associated with another Splash episode later on.

This figure also shows that the Splash population is generally older than its high-$\upalpha$ disc counterparts. 
This result supports our previous discussion of the differences in ages between these two Milky Way populations, as inferred from the differences in chemical abundances. 
Interestingly, the peak of star formation in the high-$\upalpha$ disc usually occurs later than in the Splash, suggesting that some mergers may induce starbursts in the disc.

Finally, we note that the simulated galaxies tend to display a high number of retrograde stars, in some case much higher than what is found in the APOGEE data for the MW. 
That is also the case for G44 (a MW-MA galaxy) shown in Fig.~\ref{mg-lz_artemis} and, to a lesser extent, for G34 (MW-GES). 
This result is likely related to the much higher fractions of accreted stars obtained in {\texttt ARTEMIS} than the value estimated for the MW, as described in \citet{dillamore2023b}. 
However, as also discussed by these authors, the simulated galaxies that experienced a GE/S-like event, such as G34, tend to have lower (total) accreted fractions than those without a GE/S. 
This is because galaxies without a GE/S-like event are likely to experience more major mergers at a later time \citep{dillamore2022}. 
Also, at least two galaxies in the entire {\texttt ARTEMIS} sample, G38 and G19,  have accreted fractions comparable to the MW, while G34 also provides a relatively good match (see figure 2 of \citealt{dillamore2023b}).

\section{Discussion and Conclusions}
\label{sec:discussion}

The results presented in this paper make use of observational data from APOGEE as well as the \texttt{ARTEMIS} simulations to help constrain the origin of the Splash population. 
The APOGEE data allow us to make a comparison between the Splash and the rest of the high-$\upalpha$ disc for 16 different chemical abundance ratios, as well as to study the distribution of stars in the solar neighbourhood in a way that minimises the effects of the selection function.  
These results along with the analysis of the simulations give us a clearer picture of how to tackle the problem of understanding the true origin of the Splash.  

From the results obtained on the basis of APOGEE, we find that our simple selection of the Splash agrees with previous works (e.g., \citealt{belokurov2020}) and behaves kinematically as expected, as a halo-like component, considering our selection (see Fig.~\ref{velocity}).   
We also studied the spatial distribution of the Splash by calculating the Splash fraction in different ($R,|Z|$) regions. We found that the Splash fractions in the MW are higher at smaller radial distances and at higher vertical heights, in qualitative agreement with previous observational results (e.g., \citealt{belokurov2020}) and with results from the {\texttt Auriga} simulations \citep{grand2020}. We also performed more quantitative comparisons with the Splash fractions in the {\texttt ARTEMIS} simulations, and found good agreement between measurements in the MW and the predicted values in simulated galaxies with GE/S-like events (see Figs.~\ref{confusion} and ~\ref{ConfusionSims}).

As described in Section~\ref{danny}, we compared in detail the chemical compositions of the Splash and high-$\upalpha$ disc populations.  The results, presented in Section~\ref{dannyresult}, revealed two populations that are statistically distinct from one another, as seen from the $\chi^2$ and p-values in Fig.~\ref{chemcomp}. 
Moreover, we found chemical signatures that suggest that the Splash is an older population than the high-$\upalpha$ disc, in agreement with the hypothesis that the Splash is the oldest disk population, heated up by the collision with the GE/S system.  As a consistency check, we performed another test, described in Section~\ref{dom}.  
The results presented in Section~\ref{domresult} uncover which elements have the largest contribution to the differences we see in the first test.  These are shown in Fig.~\ref{chi2}.  

We also examined the distribution of Splash and high-$\upalpha$ stars in the [X/Fe]--[Fe/H] plane for these elements (Fig.~\ref{x-fe}).  
For example, Al shows a small difference between the running median of the Splash and high-$\upalpha$ disc, at most $0.05$~dex, yet we still find that the two populations differ at a statistically significant level.  We note that the measurement of such small, yet statistically different, measurements of chemical compositions across a variety of elements is only possible because of the high-precision data from APOGEE.

Although the chemistry of the Splash is significantly different from the rest of the high-$\upalpha$ disc, as shown in Figs.~\ref{chemcomp} and \ref{chi2}, we find there is a relatively smooth variation of the $\upalpha$-abundances with respect to eccentricity (Fig.~\ref{echem_1}) and angular momentum (top panel of Fig.~\ref{mg-lz_artemis}).  
The latter negative correlation is also seen in both simulated galaxies that had a GE/S-like merger and those that had only minor mergers (bottom panel of Fig.~\ref{mg-lz_artemis}). 

Some simulated galaxies also show a discontinuity in [Mg/Fe] just at the onset of the disc spin-up, at Lz $\simeq$ 0 (see Fig.~\ref{mg-lz_artemis}).  
This behaviour is observed in other simulated galaxies, of both MW-GES and MW-MA types. This discontinuity could be associated with a burst of star formation around the time of the disc spin-up as seen, for example, in  \citet{belokurov2022}. We leave the investigation of this interesting feature for a future study.


To further identify the differences between galaxies with and without a GE/S-like event, we separated the retrograde and prograde stars in the six simulated galaxies and calculate the ratio between the two.  
For the {\it in situ} stars, this corresponds to the Splash fraction, while for the accreted stars this is the accreted retrograde fraction.   Fig.~\ref{splashfracvaccretrofrac} shows a strong positive correlation between the Splash and the accreted retrograde fractions. 
This result is similar to figures 10 and 16 from \citet{grand2020} and \citet{dillamore2022}, respectively.  However, here we chose the accreted {\it retrograde} ratio rather than the accreted ratio, because we find it correlates more strongly with the Splash fraction.  
This type of diagnostic can be used in the future to help identify  other potential contributors to the Splash.

There is, however, some scatter in this relation, as we can see from the distribution of G29, G42 and G44 in this plane. A more in-depth analysis of  the simulations, including the entire {\texttt ARTEMIS} sample of MW-mass hosts, is required to pinpoint the origin of this scatter.  
Although this is beyond the scope of the current paper, we note that multiple factors may lead to variations in the measured fractions, such as the mass ratio of MMAP to its host, the time when MMAP merges with the host, the orientation of the satellite orbital plane with respect to the disc, or the dynamical state of the disc at that time.

We have shown that multiple minor mergers are also capable of contributing to the overall Splash fraction, as well as to the accreted retrograde fraction, as long as these progenitors are on retrograde orbits. 
This is exemplified in Figs.~\ref{splashfracvaccretrofrac} and ~\ref{ConfusionSims}, where it is shown that MW-MA galaxies, i.e., without a GE/S-like event, may also contain large Splash fractions. 
The case of G44, which is an MW-MA galaxy with a higher Splash fraction than two MW-GES galaxies (G34 and G42) suggests that GE/S-like mergers are  not a necessary condition for a Splash population to exist. 
The most predictive factor for the existence of a Splash is the total mass of the accreted population that is in retrograde motion.  Figs.~\ref{fig:mergers} and \ref{fig:ages} show how Splash populations can form in a variety of merger histories.

So far in the literature, the Splash has been regarded as the result of MW's merger with the GE/S progenitor (\citealp{gallart2019, grand2020, belokurov2020, belokurov2022, xiang2022, ciuca2023, lee2023}).  
In this paper, we asked whether there is a chemo-dynamical difference between galaxies that have undergone a GE/S-like merger (MW-GES) and those that did not (MW-MA).  
While previous simulations have shown that a GE/S-mass merger results in a Splash-like population \citep{grand2020, belokurov2020}, our results prompt the question of whether a Splash component can form via more minor mergers. 
While we found that to be the case, we note that previous studies also show that Splash populations are ubiquitous in simulations. For example, \citet{dillamore2022} found that discs can also be perturbed by mergers other than with the MMAP (see their figure 14 and associated discussion).  
\citealt{grand2020} also show that Splash populations may also be produced by lower mass ($\sim 10^{8} \, {\rm M}_{\odot}$) events (see their figure 10). While these events are still classified as 'GES' in their study, their results are consistent with our findings that more minor events may generate Splash-like features. 
Furthermore, it is not inconceivable that even less massive progenitors may create a Splash, as long as they are retrograde and the conditions for impacting the disc are in place. Our results further emphasize that, in some cases, there are no strong chemo-dynamical differences between Splash populations between MW-GES and MW-MA galaxies.

We note, though, that the results presented here give only an indication of the contribution of the Splash in the solar neighbourhood. To form a full picture of the extent of the Splash, the analysis should be carried out across larger regions of a galaxy. However, while this is possible for the simulations, we need to be cautious of any selection effects that may play a part when studying the MW.

\vspace{0.1in}
Our conclusions can be summarised as follows:

\begin{itemize}
  \item In this study, we selected a sample of high-$\upalpha$ stars in the MW's solar neighbourhood ($d_\odot < 3$~kpc), which was then split on the basis of eccentricity.  Stars with $e>0.6$ were selected as the Splash population, while the rest were the remaining high-$\upalpha$ population.  
  The $(v_R, v_\phi, v_Z)$ distributions of the Splash population in our sample match the findings of previous studies.  As expected, the Splash is a low $v_\phi$ component, and has a more extended distribution in both the $v_Z$ and $v_R$ components compared to the disc.  
  \item We also examined the spatial distribution of the Splash population in the solar neighborhood, and found a larger contribution from the Splash at smaller radial ($R \simeq 5-7$~kpc) and larger heights ($|Z| \simeq 2-3$~kpc) from the Galactic centre. This is in qualitative agreement with previous analyses in the MW, and in good quantitative agreement with the spatial distribution of Splash-like features in the {\texttt{ARTEMIS}} simulations.
  
  \item Employing two different methods, we found systematic differences in the chemical abundances of the Splash population compared with those of the high-$\upalpha$ disc.  Out of 16 individual element abundances studied here, 12 show statistically significant differences. In particular, Splash stars are enhanced in [$\upalpha$/Fe] than their high $\upalpha$-disk counterparts, suggesting that they are older.  This result brings additional evidence in support of the Splash being comprised of stars belonging to the early Galactic disk, which were heated up by interaction with one or more accreted satellites.
 
  \item When we split the high-$\upalpha$ population into multiple eccentricity bins and look for correlations with abundance ratios, we find a correlation between the abundance ratios and eccentricity for most of the elements, especially for the $\upalpha$ abundances.  The chemistry of each eccentricity bin is statistically distinct from all others.  This highlights the correlation between chemistry and the kinematics in the Galaxy and how comparing kinematically selected populations can result in large chemical differences.  

  \item Using simulations, we investigated whether the merger(s) associated with the Splash leave a signature in the form of a discontinuity in the relation between chemistry and kinematics.   In particular, we focused on the [Mg/Fe]--Lz plane of the 'high-$\upalpha$' disc, where we identified a sharp decrease in [Mg/Fe] at Lz$\simeq$0. Interestingly, we find a discontinuity and a negative correlation in both MW-GES and MW-MA galaxies, with no clear difference between the two types. However, this discontinuity is not clearly seen in our MW sample.

  \item We also explored the relation between the Splash fractions and accreted retrograde fractions in the simulations and we found a strong positive correlation between the two.  We also found cases where MW-MA galaxies have higher Splash fraction than some MW-GES galaxies, which suggests that a Splash may be created by less massive mergers as long as they are on retrograde orbits.  
  
  \item While it has previously been shown that the Splash fraction correlates with accreted mass fraction \citep[e.g.,][]{dillamore2022,grand2020}, our results show that there is an even stronger correlation between the Splash fraction and the retrograde accreted fraction.  This is an important distinction, which implies that the impact of a given accretion on disc kinematics is dependent on the orbital orientation of the accreted systems.  A relatively low mass retrograde accretion can be as impactful, if not more, than a more massive accretion that happens to be prograde.

\end{itemize}


\section*{Acknowledgements}
The authors thank the anonymous referee for constructive feedback that improved the manuscript significantly, and Rob Grand for insightful discussions about the Splash.
S.S.K. acknowledges a Science and Technologies Facilities Council (STFC) doctoral studentship. 
J.G.F.-T. gratefully acknowledges the grants support provided by ANID Fondecyt Iniciaci\'on No. 11220340, ANID Fondecyt Postdoc No. 3230001 (Sponsoring researcher), from two Joint Committee ESO-Government of Chile grants under the agreements 2021 ORP 023/2021 and 2023 ORP 062/2023.
T.C.B. acknowledges partial support for this work from grant number PHY 14–30152; Physics Frontier Center/JINA Center for the Evolution of the Elements (JINA-CEE), and OISE-1927130: The International Research Network for Nuclear Astrophysics (IReNA), awarded by the US National Science Foundation.

Funding for the Sloan Digital Sky Survey IV has been provided by the Alfred P. Sloan Foundation, the U.S. Department of Energy Office of Science, and the Participating Institutions. SDSS acknowledges support and resources from the Center for High-Performance Computing at the University of Utah. The SDSS website is www.sdss4.org.

SDSS is managed by the Astrophysical Research Consortium for the Participating Institutions of the SDSS Collaboration including the Brazilian Participation Group, the Carnegie Institution for Science, Carnegie Mellon University, Center for Astrophysics | Harvard \& Smithsonian (CfA), the Chilean Participation Group, the French Participation Group, Instituto de Astrofísica de Canarias, The Johns Hopkins University, Kavli Institute for the Physics and Mathematics of the Universe (IPMU) / University of Tokyo, the Korean Participation Group, Lawrence Berkeley National Laboratory, Leibniz Institut für Astrophysik Potsdam (AIP), Max-Planck-Institut für Astronomie (MPIA Heidelberg), Max-Planck-Institut für Astrophysik (MPA Garching), Max-Planck-Institut für Extraterrestrische Physik (MPE), National Astronomical Observatories of China, New Mexico State University, New York University, University of Notre Dame, Observatório Nacional / MCTI, The Ohio State University, Pennsylvania State University, Shanghai Astronomical Observatory, United Kingdom Participation Group, Universidad Nacional Autónoma de México, University of Arizona, University of Colorado Boulder, University of Oxford, University of Portsmouth, University of Utah, University of Virginia, University of Washington, University of Wisconsin, Vanderbilt University, and Yale University.

\section*{Data Availability}
Data from the \texttt{ARTEMIS} simulations may be shared on reasonable request to the corresponding author.  All APOGEE data used in this work are in the public domain. For the purpose of open access, the author has applied a Creative Commons Attribution (CC BY) license to any Author Accepted Manuscript version arising.




\input{paper.bbl}






\bsp	
\label{lastpage}
\include{full_appendix_table}

\end{document}